\documentclass[a4paper,11pt]{article}
\usepackage[utf8]{inputenc}
\pdfoutput=1 

\usepackage{jheppub} 
\usepackage{natbib}
\usepackage[T1]{fontenc} 
\usepackage{slashed}
\usepackage{subcaption}
\usepackage{physics}
\usepackage{csquotes}
\usepackage{caption}

\usepackage{xcolor}

\title{Inflection Point Inflation in Supergravity}
\author[a]{Manuel  Drees} \author[a,b]{Wenbin Zhao}
\affiliation[a]{Bethe Center for Theoretical Physics and Physikalisches
  Institut, Universität Bonn, Nussallee 12, 53115 Bonn, Germany}
\affiliation[b]{International Centre for Theoretical Physics Asia-Pacific, University of Chinese Academy of Sciences,
100190 Beĳing, China\footnote{After
October 1, 2025.}}
\emailAdd{drees@th.physik.uni-bonn.de}
\emailAdd{wenbin.zhao@uni-bonn.de}

\abstract{In this paper, we study the inflection point inflation
  generated by a polynomial superpotential and a canonical K\"ahler
  potential under the supergravity framework, where only one chiral
  superfield is needed. We find that the special form of the scalar
  potential limits the inflationary Hubble parameter to values
  $\lesssim 10^{10}\, \textrm{GeV}$ and the inflaton mass to
  $\lesssim 10^{11} \, \textrm{GeV}$. We obtain analytic results for
  small field cases and present numerical results for large field
  ones. We find the tensor-to-scalar ratio $r<10^{-8}$ is always
  suppressed in these models, while the running of spectral index
  $\alpha\approx \mathcal{O}(-10^{-3})$ may be testable in
  next-generation CMB experiments.  We also discuss the possible
  effects of a SUSY breaking Polonyi term presented in the
  superpotential where we find a general upper bound for the SUSY
  breaking scale for a given value of the Hubble parameter.}

\begin{document}

\maketitle
\section{Introduction}

Recent observations of the cosmic microwave background (CMB) by the
Planck and BICEP/Keck experiments strongly favor an exponentially
expanding period of the early universe~\cite{Planck:2018jri,
  BICEP:2021xfz}, which is called inflation~\cite{Starobinsky:1980te,
  Guth:1980zm, Linde:1981mu, Albrecht:1982wi}. The simplest
implementation, slow roll inflation, needs a single scalar
``inflaton'' field $\phi$ rolling over a flat region in the
potential. The flatness of the potential can be parameterized by the
following slow roll parameters\footnote{We set the reduced Planck Mass
  $M_{\textrm{pl}}=\sqrt{\frac{1}{8\pi G}}\approx 2.4 \times 10^{18}
  \,\textrm{GeV}=1$ throughout this paper.}~\cite{Lyth:2009imm}:
\begin{equation}
    \epsilon = \frac{1}{2}\left( \frac{V'}{V}\right)^2  \,, \quad 
    \eta = \frac{V''}{V} \,, \quad \xi^2 = \frac{V'V'''}{V^2} \,,
\end{equation}
where a prime ($'$) denotes the derivative with respect to
$\phi$. They are related to the CMB
observations~\cite{Planck:2018jri,BICEP:2021xfz}:\footnote{We will
  comment on the very recent ACT determination \cite{ACT:2025fju,
    ACT:2025tim} of $n_s$ in Sec.~3.1, following
  eq.(\ref{eq:SUGRA_results}).}
\begin{equation}    \label{eq:PlanckResults}
    \begin{split}
        P_\zeta &= \frac{V}{24\pi^2 \epsilon}=(2.1\pm 0.1) \times 10^{-9} \,;\\ 
        n_s &= 1-6\epsilon+2\eta = 0.9659 \pm  0.0040 \,;\\ 
        r_{0.05} &= 16 \epsilon <0.036 \,;\\
        \alpha &= 16 \epsilon \eta -24 \epsilon^2-2\xi^2 = -0.0041 \pm 0.0067\,.
    \end{split}
\end{equation}
Here $P_\zeta$ is the power spectrum of Gaussian curvature
perturbations, $n_s$ is the spectral index, $r$ is the
tensor-to-scalar ratio and $\alpha$ is the running of the spectral
index.

Power law potentials $V\propto\phi^n$ offer one of the most elegant
and simplest realizations of inflation. However, such potentials with
$n>1$ have been ruled out by the CMB observations, due to the large
tensor-to-scalar ratio predicted by these models. Given this fact, a
broad program has been pursued to investigate alternative models for
inflation. Among them are the so-called inflection point inflation
models, where different $\phi^n$ terms collectively provide an
inflection point in the potential and accommodate inflation around
it. It has been realized using flat directions of the Minimally
Supersymmetric Standard Model~\cite{Allahverdi:2006iq,
  Allahverdi:2006cx, Allahverdi:2006we, Enqvist:2010vd}, in the
context of Higgs inflation~\cite{Okada:2016ssd, Okada:2017cvy} and
warped $D-$brane inflation models~\cite{Baumann:2007np,
  Baumann:2007ah}. As a generic model, it has been well studied both
during and after inflation~\cite{Hotchkiss:2011am, Chang:2013cba,
  Martin:2013tda, Martin:2013nzq, Choi:2016eif, Dimopoulos:2017xox,
  Musoke:2017frr, Drees:2021wgd, Drees:2022aea, Drees:2024hok,
  Bernal:2024ykj, Drees:2025iue}.

Supergravity emerges as the low energy limit of superstring theory,
which is probably our best bet for a perturbative understanding of
quantum gravity. Moreover, supersymmetry protects the (super)potential
from loop corrections, which otherwise can easily spoil the required
flatness of the inflaton potential.\footnote{In large field models,
  quantum corrections to the K\"ahler metric can still be dangerous.}
In this paper we therefore consider inflation models in the
supergravity (SUGRA) framework~\cite{Nanopoulos:1982bv, Adams:1992bn,
  Copeland:1994vg, Kumekawa:1994gx, Izawa:1996dv,
  Panagiotakopoulos:1997qd, Linde:1997sj, Kawasaki:2000yn,
  Binetruy:2004hh, Ellis:2013xoa, Ellis:2013nxa, Kallosh:2013yoa,
  Carrasco:2015pla}. Supergravity models with inflection point
inflation have previously been constructed with canonical K\"ahler
potential in~\cite{Holman:1984yj, Ross:1995dq, Mazumdar:2011ih}, with
logarithmic K\"ahler potential in~\cite{Linde:2007jn, Conlon:2008cj,
  Maharana:2015saa, Gao:2015yha, Pallis:2023aom}, through a stabilizer
field in~\cite{Gao:2016xfv} and with extended supergravity
in~\cite{Aldabergenov:2020bpt}.

In this paper, we focus on a set--up with a single chiral
superfield. A no-go theorem shows that a single chiral superfield with
logarithmic K\"ahler potential can not support
inflation~\cite{Badziak:2008yg, Ben-Dayan:2008fhy, Covi:2008cn},
unless a modification of K\"ahler potential is
made~\cite{Ketov:2014qha}. Hence, we choose to work with canonical
K\"ahler potential. For simplicity, we consider a polynomial
superpotential with three terms. This mirrors non--supersymmetric
inflection point inflation with renormalizable potential
\cite{Drees:2021wgd}, and allows us to control the position of the
inflection point explicitly. Flexibility and control are among the
novel features of our model.

There is a price we have to pay. Like many other inflection point
models, parameters in this model are fine-tuned to support inflation
and fit the CMB observables. There is no symmetry argument indicating
these parameters should take specific values.  In this sense, our
model belongs to the category where inflation happens accidentally in
the parameter space~\cite{Linde:2007jn}. Some progress has been made
to reduce the fine-tuning, as in ~\cite{Enqvist:2010vd,
  Hotchkiss:2011am, Dimopoulos:2017xox}. It is beyond the scope of
this paper to address this problem.

SUSY must be broken since no superpartner of a Standard Model particle
has as yet been detected. Once SUSY breaking terms are included, they
may modify the form of the inflaton potential, thereby breaking the
slow-roll conditions. These corrections will be under control if the
SUSY breaking scale is not too large. We investigate this issue for
the simple case where SUSY is broken by the Polonyi
model~\cite{Polonyi:1977pj}, which can play an important role in
cosmology. In particular, the late time decay of Polonyi particles may
adversely impact Big Bang nucleosynthesis (BBN); this is known as the
``Polonyi problem''~\cite{Ellis:1986zt, Coughlan:1983ci}. In F-term
inflation models, this leads to a stringent constraint on the Hubble
parameter during inflation~\cite{Ibe:2006am}. For D-term inflation, it
has been studied in~\cite{Addazi:2017ulg, Aldabergenov:2017bjt,
  Aldabergenov:2017hvp, Abe:2018rnu}. Here we focus on the impact of
the Polonyi superpotential on the inflation potential. It has been
shown that no-scale Starobinsky-like inflation is only compatible with
a Polonyi sector if the SUSY breaking scale is smaller than
$10^3\, \textrm{TeV}$~\cite{Romao:2017uwa}. In this paper, we analyze
two limiting cases by comparing the relative strengths of the
inflation sector and the SUSY breaking sector. They lead to very
different bounds on the SUSY breaking scale, which can be quite large
in our model. This is another new result presented in this paper,
which defuses the Polonyi problem, since the lifetime of Polonyi
particles scales with the inverse third power of their mass; they will
thus decay before the onset of BBN if their mass exceeds some tens of
TeV.

In this paper we present a systematic study allowing us to scan over
the possible parameter space based on the position of the inflection
point. We will focus on the similarities and differences between the
previously studied non-supersymmetric, renormalizable models and SUGRA
models. To be more precise, we consider a SUSY-preserving inflaton
field $\phi$, which accommodates a near inflection point in the scalar
potential at $\phi = \phi_0$. We find that the Hubble parameter and
the inflaton mass both increase with $\phi_0$ as long as $\phi_0
<1$. They reach a maximum around $\phi_0 \sim 1$, then start to
decrease because of the exponential factor in the potential. Thus, in
such a model, the Hubble parameter can not exceed
$\mathcal{O}(10^{10})\ \textrm{GeV}$. The tensor-to-scalar ratio $r$
is much smaller than the current upper bound. The model also predicts
a near constant running $\alpha \sim -0.003$, a unique feature that
the next generation of observations might be able to
test~\cite{Kohri:2013mxa, CMB-S4:2016ple, Munoz:2016owz,
  Modak:2021zgb, Easther:2021rdg, Bahr-Kalus:2022prj}.

The remainder of this paper is organized as follows. In sec.~2 we
recap basic results obtained in the non-supersymmetric, renormalizable
version of the model and argue why it is interesting to consider this
scenario in the SUGRA framework. In sec.~3 we present our model's
analytic and numerical results. We further discuss the SUSY breaking
effects in our model. In sec.~4 we summarize the work and draw some
conclusions.

\section{Non-supersymmetric Inflection Point Model and Beyond}       
\subsection{Potential setup and CMB observables}  

We start from the most general renormalizable potential of a single
real scalar inflaton field $\phi$ as a non-supersymmetric example:
\begin{equation}     \label{eq:Renormalphi4}\,
  V(\phi)=b\phi^2+c\phi^3+d\phi^4\,,
\end{equation}
where we have removed the constant and linear terms, so that the
minimum of the potential is defined to be at $\phi = 0$ with vanishing
vacuum energy\footnote{A tiny cosmological constant can be added.}.
Requiring an inflection point, $V'(\phi_0)=V''(\phi_0)=0$, then leads
to:
\begin{equation}
    b=\frac{9c^2}{32d}\,, \quad \phi_0 = -\frac{3c}{8d}\,.
\end{equation}
CMB observations indicate the potential is not exactly flat, but
rather concave. One way to realize this is to introduce a small
deviation from the inflection point conditions in the cubic term. For
our purpose it is more convenient to write the coefficients in terms
of the inflection point position $\phi_0$. The modified potential then
reads:
\begin{equation} \label{eq:V1}
  V(\phi) = d \left( \phi^4 - \frac{8}{3} \phi_0 (1 - \beta) \phi^3
    +2 \phi_0^2 \phi^2 \right)\,.
\end{equation}
There are three free parameters in the potential, $d,\phi_0$ and
$\beta$. In eq.(\ref{eq:V1}), $d$ determines the overall normalization
of the potential, which can be matched to the power spectrum of
curvature perturbation $P_\zeta$ once the other parameters are
fixed. The other two parameters govern the shape the potential and
hence determine the CMB observables, such as the number of e-folds of
inflation after the CMB pivot scale left the horizon,
$N_\textrm{cmb}$, and the spectral index $n_s$. In a ``small field''
set--up, i.e. for sub-Planckian field values, $\phi_0 <1$, fixing
$N_\textrm{cmb}=65$ and eq.\eqref{eq:PlanckResults}
require~\cite{Drees:2021wgd}:
\begin{equation}
\begin{split}  \label{eq:reinflationsol}
  d&= 6.61 \times 10^{-16} \phi_0^2\,; \\
 \beta &= 9.73 \times 10^{-7} \phi_0^4\,.
\end{split}
\end{equation}
This leads to the following predictions:
\begin{equation} \label{eq:phi4results}
    \begin{split}
      b&= 1.3\times 10^{-15}\,\phi_0^4\,; \quad
      c= 1.8 \times 10^{-15}\,\phi_0^3\,;\\ \indent
      H_{\textrm{inf}}& = 8.6 \times 10^{-9}\, \phi_0^3\,; \quad
      m_\phi = 5.2 \times 10^{-8}\,\phi_0^2\,;\\ \indent
       r &= 7.1 \times 10^{-9}\, \phi_0^6\,; \quad \alpha=-1.4\times10^{-3}\,.
    \end{split}
\end{equation}
The same ansatz for the potential can also describe ``large field''
scenarios, where $\phi_0>1$. However, in this case no analytical
treatment is known. Numerical studies showed that this model can cover
the whole allowed parameter space in the $n_s-r$ plane and may even
lead to double eternal inflation~\cite{Drees:2022aea}.

\subsection{Possible Realization in SUGRA and Associated Problem}

In SUGRA\footnote{For a detailed introduction, please
  consult~\cite{Freedman:2012zz}.} the scalar potential is generated
by the real K\"ahler potential $K$ and the holomorphic Superpotential
$W$. Both are supposed to be functions of the complex field $\Phi$,
which is the scalar component of a chiral superfield. The K\"ahler
potential also determines the kinetic term of the scalar field (in the
convention of ref.~\cite{Freedman:2012zz}):
\begin{equation}
  \mathcal{L}_\textrm{kin} = - \frac {\partial^2 K}
  {\partial \Phi \partial \Phi^*} \partial_\mu \Phi \partial^\mu \Phi^* \,.
\end{equation}
A canonically normalized field thus requires $K = \Phi \Phi^*$ and we
will use this K\"ahler potential throughout the paper. We can further
define the K\"ahler covariant derivative as:
\begin{equation}\label{eq:Kderivative}
  D_\Phi W = \frac {\partial W} {\partial \Phi}
  + \frac {\partial  K} {\partial \Phi} W \, =
   \frac {\partial W} {\partial \Phi} + \Phi^* W\,.
\end{equation}
The $F-$term contribution to the scalar potential then reads:
\begin{equation} \label{eq:mSUGRApotential}
  V(\phi) = e^{K} [ \abs{ D_\phi W }^2 -3 \abs{W}^2] \,.
\end{equation}
Since we assume the inflaton to be a gauge singlet, there is no $D-$term
contribution.

For simplicity we choose the superpotential to be a polynomial
function of the complex field $\Phi$:
\begin{equation} \label{eq:W1}
    W(\Phi) = B\Phi^2+C\Phi^3+D\Phi^4 \,,
\end{equation}
where we neglect the constant and linear terms as in the
renormalizable case. The ansatz (\ref{eq:W1}) ensures that
$V(0) = W(0) = 0$, i.e. the potential has a supersymmetric stationary
point at $\Phi = 0$. The three coefficients $B,C,D$ are chosen to be
real. The complex $\Phi$ can in general be written as
$\Phi = (\phi + i \chi)/\sqrt{2}$, where $\phi$ and $\chi$ are real
fields. We want to identify $\phi$ with the inflaton field.  By
choosing the coefficients in eq.(\ref{eq:W1}) to be real we make sure
that the potential $V(\phi,\chi)$ depends only on even powers of
$\chi$. Moreover, we make sure that
$\partial^2 V(\phi,\chi) / \partial \chi^2 > 0$ for $\chi = 0$ and
$\phi \in [0, \phi_0]$. We therefore can assume that $\chi = 0$
throughout, so that we can ignore this imaginary component when
computing the inflationary dynamics. It is not hard to write down the
scalar potential $V(\phi) \equiv V(\phi,\chi=0)$ in this setup. We
first notice that the two terms in eq.\eqref{eq:mSUGRApotential} read:
\begin{equation}
\begin{split}
  \abs{ D_\phi W }^2 &= \frac{1}{64} \phi ^2 \left[2 \sqrt{2} B \left(\phi ^2
      +4\right)+\phi  \left(2 C \left(\phi ^2+6\right)+\sqrt{2} D \phi
      \left(\phi ^2+8\right)\right)\right]^2 \,,\\
  3\abs{ W }^2 &= \frac{3}{16} \phi ^4 \left[2 B+\phi  \left(\sqrt{2} C
      +D \phi \right)\right]^2 \,.
\end{split}
\end{equation}
The combination of them leads to the full potential:
\begin{equation}  \label{eq:full_potential}
 \begin{split}
  V(\phi)= {\rm e}^{\frac{\phi^2}{2}} \Biggl[ 2 B^2 \phi ^2
  + 3 \sqrt{2} B C \phi^3 + \frac {1}{4} \phi^4 \left( B^2 + 16 B D+9 C^2\right)
  + \frac {1}{2} \phi^5 \left( \sqrt{2} B C + 6 \sqrt{2} C D \right)
  \\ \indent
  + \frac {1}{8} \phi^6 \left( B^2 + 6 B D + 3 C^2 + 16 D^2\right)
  + \frac {1}{8} \phi^7 \left( \sqrt{2} B C + 4 \sqrt{2} C D \right)
  \\ \indent
  + \frac {1} {16} \phi^8 \left( 2 B D + C^2 + 5 D^2\right)
  + \frac {C D \phi^9} {8 \sqrt{2}} + \frac{D^2 \phi^{10}} {32}\Biggr]\,,
\end{split}
\end{equation}
Note that only plus signs appear in the above
  potential, even though we start from eq.\eqref{eq:mSUGRApotential}
  with a minus sign in front of $\abs{W}^2$. This comes from the fact
  that the mixing term in $\abs{ D_\phi W }^2$,
  $\Phi W^* \partial W/ \partial \Phi + c.c.$, gives the same
  combinations of $B,C,D$ and powers of $\phi$ as in $-3\abs{W}^2$,
  but with larger coefficients. Of course, this by itself doesn't mean
  that the potential is always positive, as the parameters $B,C,D$ as
  well as the field $\phi$ can be either positive or negative.

We first try to match to the non-supersymmetric case by expanding $V$ up
to the fourth order:
\begin{equation}  \label{eq:SUGRAphi4}
 V(\phi) \approx 2 B^2 \phi^2 + 3 \sqrt{2} B C \phi^3 + 
 \frac{1}{4} (5 B^2 + 9 C^2 + 16 B D) \phi^4\,.
\end{equation}
Naively these are the leading terms if $\phi \ll 1$. Matching
eq.(\ref{eq:SUGRAphi4}) to eq.(\ref{eq:Renormalphi4}) and using
eqs.(\ref{eq:reinflationsol}) and (\ref{eq:phi4results}) implies:
\begin{equation}  \label{eq:matching}
\begin{split}
    B &= 2.571 \times 10^{-8} \phi_0^2\,,\\
    C &= 1.615 \times 10^{-8} \phi_0 \times (1-\beta)\,,\\
    D &= 7.210 \times 10^{-10} - 8.034 \times 10^{-9} \phi_0^2
    - 5.7 \times 10^{-8}(2\beta-\beta^2)\,.
\end{split}
\end{equation}
Hence for $\phi_0 \ll 1$ the coefficients of the superpotential would
have to scale as
$B \propto \phi_0^2,\ C \propto \phi_0^1,\ D \propto \phi_0^0$. The
problem emerges when we substitute these matching conditions back into
the full potential eq.\eqref{eq:full_potential}. Writing the latter
as:
\begin{equation}
    V(\phi) = e^{\frac{1}{2}\phi^2} \sum_{n=2}^{10} a_n \phi^n\,,
\end{equation}
the coefficients $a_n$ have the following scaling behavior for
$\phi_0 \ll 1$:
\begin{equation}
\begin{split}
  & a_2 \propto \phi_0^4; \quad a_3 \propto \phi_0^3; \quad
  a_4 \propto \phi_0^2; \quad a_5 \propto \phi_0^1; \quad
  a_6 \propto \phi_0^0;\\ \indent
  &a_7 \propto \phi_0^1; \quad a_8 \propto \phi_0^0; \quad
  a_9 \propto \phi_0^1;\quad a_{10} \propto \phi_0^0.
  \end{split}
\end{equation}
Thus, when we evaluate the value of the potential and its first and
second derivatives at $\phi \sim \phi_0$, the first five terms (from
$\phi^2$ to $\phi^6$) would contribute with comparable
magnitude. Therefore the terms $\propto \phi^5$ and $\phi^6$ can
easily spoil the flatness of the potential, i.e. the expansion only to
order $\phi^4$ is not self--consistent. For this reason, it is
necessary to consider an inflection point model in the full potential
eq.\eqref{eq:full_potential} (or at least keeping terms up to $\phi^6$
in the small field case) rather than trying to directly match the
renormalizable, non-supersymmetric potential.

\section{Inflection Point Model in SUGRA}
\subsection{Analytic Analysis of the Model}

Requiring that the full potential eq.\eqref{eq:full_potential} has an
inflection point at $\phi = \phi_0$, i.e. $V'(\phi_0)=V''(\phi_0)=0$,
leads to the following solutions\footnote{Since we are dealing with
  high-order polynomial equations there often are several
  solutions. However, we find that the others usually have $V < 0$ at
  the minimum, leading to a very large and negative cosmological
  constant.}:
\begin{equation} \label{eq:solution_BC}
    \begin{split}
      B =& D \frac {\phi_0^2 \left( 1152 + 480 \phi_0^2 + 72 \phi_0^4
          + 12 \phi_0^6 + \phi_0^8 \right)}
      {2 \left( 192 + 96 \phi_0^2 + 4 \phi_0^6 + \phi_0^8 \right)}\,;\\
      C =& -D \frac {\sqrt{2} \phi_0  \left( 384 + 192 \phi_0^2 + 24 \phi_0^4
          + 8 \phi_0^6 + \phi_0^8 \right)}
      {192 + 96 \phi_0^2 + 4 \phi_0^6 + \phi_0^8 }\,.
    \end{split}
\end{equation}
The inflection point condition allows us to bypass
  the $\eta$ problem for the canonical K\"ahler
  potential~\cite{Copeland:1994vg,Stewart:1994ts}, which assumes that
  a single term in the second derivatives $V''$, arising from the
  exponential factor, dominates over the other possible ones. For a
  scalar potential accommodating an inflection point, this assumption
  doesn't hold, and a flat potential appears only when we go beyond
  it. As already noted in the Introduction, this requires tuning of a
  priori unrelated parameters, in particular of the ratio $B/C$, as
  described by eqs.(\ref{eq:solution_BC}).

For $\phi_0 \ll 1$ we can approximate the above solutions by their
leading order results:
\begin{equation*}
 B \approx D\times 3 \phi_0^2\,, \quad C \approx D \times(-2\sqrt{2}\phi_0)\,.
\end{equation*}
The full potential at $\phi \leq \phi_0$ can be further simplified if
we only include terms up to $\phi_0^6$:
\begin{equation}\label{eq:Sim_potential}
 V(\phi) \approx 2 D^2 \phi^2 ( \phi^2 - 3\phi \phi_0 + 3\phi_0^2)^2\,.
\end{equation}
This expansion now {\em is} self-consistent, i.e. the higher order
terms, starting at ${\cal O}(\phi^7)$, are indeed suppressed.  At the
inflection point $\phi_0$, the potential reads:
\begin{equation}\label{eq:potential_value}
    V(\phi_0) \approx 2D^2 \phi_0^6\,.
\end{equation}
The potential~\eqref{eq:Sim_potential} has a
  minimum at $\phi=0$, with $V(0) = 0$. The algebraic equation
  $D_\phi W\pm \sqrt{3}W=0$ has no real root besides $\phi=0$ when
  using eq.\eqref{eq:solution_BC}, which can be verified by checking
  the discriminant of the equation with additional combination of
  coefficients~\cite{Rees:1922}. Hence the full potential of
  eq.(\ref{eq:full_potential}) is positive semi-definite,
  i.e. $V(\phi) \geq 0 \ \forall \phi$. [This is immediately obvious
  for the simplified potential of eq.(\ref{eq:Sim_potential}).] As
in the non-supersymmetric case, in the small field scenario
$\phi_0 \ll 1$ we can get semi-analytic results for inflationary
observables by expanding the slow-roll parameter around the inflection
point via the ansatz $\phi = \phi_0 (1-\delta\phi)$, and adding a
deviation from the strict inflection point condition via
$B \to B +D \times \delta B$:
\begin{equation} \label{eq:srp}
 \begin{split}
   \epsilon &= \frac{1}{2} \left( \frac {V'} {V} \right)^2
   = \frac{1}{2} \left( \frac {2\delta B + 6 \phi_0^2 \delta\phi^2}
     {\phi_0^3} \right)^2\,,\\
   \eta &= \frac{V''}{V} = \frac {10.5 \, \delta B -24\,\delta \phi}
   {2\phi_0^2}\,,\\ 
   \xi^2 &= \frac{V'V'''}{V^2} = \frac{24  \delta B + 72 \phi_0^2 \delta \phi^2}
   {\phi_0^6}\,.
    \end{split}
\end{equation}
We only keep terms up to linear order in $\delta B$ and quadratic in
$\delta \phi$; this will turn out to be sufficient.

We are now ready to discuss which $\delta \phi$ and $\delta B$
reproduce the CMB observations. In our model the duration of inflation
is controlled by $\eta$ since $\epsilon \ll |\eta|$. The beginning (when
the pivot scale $k_* = 0.05 \,\textrm{Mpc}^{-1}$ crossed out of the
horizon) and end of observable inflation are given by:
\begin{equation}
    \eta_\textrm{cmb} = \frac{n_s-1}{2}\,, \quad \eta_\textrm{end} = -1\,,
\end{equation}
Solving the second eq.(\ref{eq:srp}) for $\delta \phi$, we get:
\begin{equation}   \label{eq:deltaphi_deter}
  \delta \phi = -\frac {2\phi_0^2 \eta - 10.5 \delta B} {24} \,.
\end{equation}
The number of e-folds $N_\textrm{cmb}$ is given by:
\begin{equation} \label{eq:Ncmb}
\begin{split}
  N_\textrm{cmb} &= \int_{\phi_\textrm{cmb}}^{\phi_\textrm{end}} \frac {1}
  {\sqrt{2\epsilon}} d\phi\\ \indent
  & = \int_{\phi_\textrm{cmb}}^{\phi_\textrm{end}} \frac{\phi_0^3}
  {2\delta B + 6\phi_0^2 \delta\phi^2} d\phi\\ \indent
  & = -\int_{\delta\phi_\textrm{cmb}}^{\delta\phi_\textrm{end}} \frac{\phi_0^4}
  {2\delta B + 6\phi_0^2 \delta\phi^2} d\delta\phi\\ \indent
  & = \frac {\phi_0^2} { 6\sqrt{\beta}} \left[ \arctan \left(
      { \frac{\delta \phi_\textrm{end} } { \sqrt{\beta} } } \right)
    - \arctan \left( { \frac{\delta \phi_\textrm{cmb}}
        {\sqrt{\beta}}}\right)\right]\,.
\end{split}
\end{equation}
In the last step we have switched from the absolute deviation
$\delta B$ to the relative one
$\beta = D\times \delta B/B = \delta B/(3\phi_0^2)$ in order to factor
out $\phi_0^2$ in the denominator; except for the first line,
eq.(\ref{eq:Ncmb}) also only holds for $\phi_0 \ll 1$.  Recall that
the arc-tangent functions can be at most $\pi/2$.  Numerically,
requiring $N_\textrm{cmb}=50$ yields
$\beta = 2.7 \times 10^{-5} \phi_0^4$ and
$\delta B = 8.2 \times 10^{-5} \phi_0^6 $. From the second
eq.(\ref{eq:PlanckResults}) and remembering $|\epsilon| \ll |\eta|$ we
see that $|\eta| \geq 0.015$ when and after CMB scales crossed out of
the horizon. Hence we can neglect the second term in
eq.\eqref{eq:deltaphi_deter} for $\phi_0 \ll 1$, and determine the inflation period by:
\begin{equation}
    \delta \phi = -\frac{\phi_0^2}{12}\eta \,.
\end{equation}
Inserting this into the last line of eq.(\ref{eq:Ncmb}) yields
\begin{equation}
  N_\textrm{cmb} = \frac {\phi_0^2} { 6 \sqrt{\beta}} \left[
    \arctan\left( { \frac {-\phi_0^2 \eta_\textrm{end}}
        {12\sqrt{\beta}}} \right)
    - \arctan\left( { \frac{-\phi_0^2 \eta_\textrm{cmb}}
        {12\sqrt{\beta}}}\right)\right]\,,
\end{equation}
which can be solved numerically. For example, setting
$n_s=0.9659\,, N_\textrm{cmb}=45$ would result in
$\delta B =6.1\times10^{-5}\phi_0^6$.

Having fixed $\delta B$ and the initial $\delta \phi$, the overall
scale of inflation, and hence $D$, is determined by the power of
Gaussian curvature perturbations. To this end we first calculate
$\epsilon$ at $\phi_{\textrm{cmb}}$:
\begin{equation}
    \epsilon_\textrm{cmb}=8.88\,\times10^{-9} \phi_0^6\,,
\end{equation}
Using eqs.(\ref{eq:PlanckResults}) and (\ref{eq:potential_value}) we
find that the normalization factor is independent of $\phi_0$:
\begin{equation}
  D^{-2} = \frac {2 \phi_0^6} {P_\zeta 24 \pi^2 \epsilon_\textrm{cmb}}
  = 4.52 \times 10^{14}\,.
\end{equation}
It is then straightforward to evaluate the value of the Hubble
parameter during inflation and the physical mass of the inflaton after
inflation:
\begin{equation}
  H_\textrm{inf} = \sqrt { \frac {V(\phi_0)} {3} }
  = 3.84 \times 10^{-8} \phi_0^3 \,,\quad
  m_\phi= \sqrt{4B^2} = 2.81 \times 10^{-7} \phi_0^2\,.
\end{equation}
The running of the spectral index $\alpha$ can also be determined:
\begin{equation}
  \alpha = 16 \epsilon \eta - 24 \epsilon^2 - 2\xi^2
  \approx -2\xi^2 = -2\frac{24  \delta B + 72 \phi_0^2 \delta \phi^2}
  {\phi_0^6} \approx -0.0030\,.
\end{equation}
This running of the spectral index is a feature of our model, since it
does not depend on $\phi_0$.

However, our numerical results do depend on $N_\textrm{cmb}$. Taking
as second example the rather large value $N_\textrm{cmb}=65$ while
keeping $n_s=0.9659$, we get:
\begin{equation}\label{eq:SUGRA_results}
\begin{split}
\delta B = 2.60 \times10^{-5}\phi_0^6 \,&,\quad 
\epsilon_\textrm{cmb} = 1.59 \times 10^{-9} \phi_0^6\,,\\  
D^{-2} = 2.53 \times 10^{15}\,&, \quad
H_\textrm{inf} = 1.62 \times 10^{-8} \phi_0^3 \,,\\
m_\phi =1.19 \times 10^{-7} \phi_0^2 \,&,\quad \alpha = -0.0015 \,.
\end{split}
\end{equation}
The tensor-to-scalar ratio $r = 16 \epsilon$ is always too small to
reach the sensitivity of any currently conceivable
observation. However, the S4CMB experiment, together with small-scale
structure information (e.g. on the Lyman-$\alpha$ forest) could
achieve $10^{-3}$ sensitivity for $\alpha$, which would test our
model~\cite{Kohri:2013mxa, CMB-S4:2016ple, Munoz:2016owz,
  Modak:2021zgb, Easther:2021rdg,Bahr-Kalus:2022prj}.

Recently, the ACT experiment measured a slightly larger spectral
index when combining their results with DESI and Planck data, $n_s = 0.9743 \pm 0.0034$~\cite{DESI:2024mwx, ACT:2025fju,
  ACT:2025tim}. Using this central value gives us slightly different
predictions. Taking $N_\textrm{cmb}=45$ as an example, we have:
\begin{equation}
\begin{split}
\delta B = 6.88 \times10^{-5}\phi_0^6 \,&,\quad 
\epsilon_\textrm{cmb} = 1.05 \times 10^{-8} \phi_0^6\,,\\  
D^{-2} = 3.85 \times 10^{14}\,&, \quad
H_\textrm{inf} = 4.16 \times 10^{-8} \phi_0^3 \,,\\
m_\phi =3.06 \times 10^{-7} \phi_0^2 \,&,\quad \alpha = -0.0035 \,.
\end{split}
\end{equation}
while taking $N_\textrm{cmb}=65$, we have:
\begin{equation}\label{eq:SUGRA_results_ACT}
\begin{split}
\delta B = 2.83 \times10^{-5}\phi_0^6 \,&,\quad 
\epsilon_\textrm{cmb} = 2.02 \times 10^{-9} \phi_0^6\,,\\  
D^{-2} = 1.99 \times 10^{15}\,&, \quad
H_\textrm{inf} = 1.83 \times 10^{-8} \phi_0^3 \,,\\
m_\phi =1.34 \times 10^{-7} \phi_0^2 \,&,\quad \alpha = -0.0015 \,.
\end{split}
\end{equation}
One can see a shift in the spectral index leaves a subtle change on
inflationary observables. The larger value of the spectral index from
ACT would prolong the duration of inflation in the field space. Hence
for a fixed number of e-folds $N_\textrm{cmb}$, ACT results will
predict larger $\epsilon_\textrm{cmb}$, larger Hubble scale value
$H_\textrm{inf}$ and inflaton mass $m_\phi$. This effect can be
clearly seen by comparing the results in eq.\eqref{eq:SUGRA_results}
and \eqref{eq:SUGRA_results_ACT}. However, the differences between the
model parameters derived from PLANCK and ACT observations with fixed
$N_{\rm CMB}$ are much smaller than the variations of these parameters
when $N_{\rm CMB}$ is varied between $45$ and $65$. We will therefore
continue to use the PLANCK derived parameter values and estimate their
uncertainty by considering these two values of $N_{\rm CMB}$.

Before moving to the next part, we would like to briefly mention the
possibility of Ultra-slow-roll (USR) in our model, as inflection point
inflation is a well-known example to study the USR process. To have a
USR period, we need the kinetic energy of the inflation field to be
much larger than its slow-roll kinetic
energy~\cite{Dimopoulos:2017ged}. In order to have a stable USR, the
second order derivative of the potential has to be positive:
$V''>0$~\cite{Pattison:2018bct}. The first condition is up to the
initial conditions of the inflaton field dynamics. In our calculation,
we have assumed the evolution starts from the slow-roll phase. This
assumption can be relaxed to incorporate USR. The second condition is
stricter. Throughout the observable inflation, our potential features
a negative slow roll parameter $\eta$, i.e., $V''<0$.  This excludes a
stable USR phase.

By comparing with the predictions of the renormalizable model given in
eqs.\eqref{eq:phi4results}, we see that $H_\textrm{inf}$ and $m_\phi$
scale with $\phi_0$ in the same way in both cases, with roughly a
factor $2$ difference in coefficients. Even though the SUGRA potential
is more complicated and scales as $\phi^6$ in the simplest limit, they
thus make very similar predictions for the inflationary
observables. In particular, the overall coefficient $D$ of the sixth
order SUGRA potential is independent of $\phi_0$, while in the
non-SUSY case the coefficient of the quartic potential
$d \propto \phi_0^2$; hence in both cases
$V(\phi \simeq \phi_0) \propto \phi_0^6$.

We also note that the curvature of the potential is negative for an
extended range of $\phi$ below $\phi_0$. In the non-SUSY case, this
holds for $\phi/\phi_0 \in [1/3, 1]$, independently of the value of
$\phi_0$; in the SUGRA case with $\phi_0 \ll 1$, this region extends to
$\phi/\phi_0 \in [1/4,1]$. The minimum of the curvature occurs at
$\phi \simeq 0.54 \phi_0$, closer to the origin than in the non-SUSY
case, with a value just below $-0.23~m_\phi^2$, smaller in magnitude
than in the non-SUSY case. The latter reduces the tachyonic
instability while the former increases it. Hence, we expect the
non-perturbative effects after inflation studied
in~\cite{Drees:2025iue} for the non-SUSY case would be similar in the
SUGRA version of the model.

When $\phi_0$ is larger than unity, it is hard to make a comprehensive
analytic analysis, but we can still understand the model qualitatively.
To this end we first formally rewrite the potential as:
\begin{equation}
    V(\phi) = e^{\phi^2/2}P(\phi)\,.
\end{equation}
The slow roll parameter then scales like
$\eta \propto \phi^2 \times f(\delta \phi)$ with $ f(0) \approx
0$. The duration of inflation is still controlled by
$\delta \eta = \eta_\textrm{cmb} - \eta_\textrm{end} \approx 1$, the
larger $\phi_0$, the smaller $\delta \phi$ should be. The resulting
decrease in the integration range in eq.\eqref{eq:Ncmb} has to be
compensated by increasing the integrand, by a reduction of
$\epsilon$. Since the ratio between the potential $V(\phi_0)$ and
$\epsilon$ is fixed by the power of curvature perturbations, an
increase of $\phi_0$ would eventually lead to a decrease of the
potential and hence of the Hubble parameter. Moreover, the potential
near $\phi_0$ is increased by the exponential factor, which becomes
unity at the origin. Hence, the mass of inflaton should be suppressed
by $e^{-\frac{1}{2}\phi_0^2}$.

As a brief summary, by fixing $n_s$ and $N_\textrm{cmb}$, we find the
inflection point model would always give a tiny tensor-to-scalar ratio
$r$ and a constant running of spectral index $\alpha$. The Hubble
scale of inflation first increases with increasing inflection point
position $\phi_0$, and then decreases once $\phi_0$ exceeds $1$. The
inflaton mass follows the same pattern but drops much faster in the
second phase. In the next section, we confirm these expectations by
showing some numerical results.

\subsection{Numerical Results of the Model}

In this section, we present our numerical results. We introduce three
steps to scan the allowed parameter space of our model, following the
same spirit as our analytic treatment:

\begin{itemize}
  
\item We choose $\phi_0$ as a free parameter and solve the inflection
  point equations $V'=V''=0$ to find corresponding values of $B/D$ and
  $C/D$. We pick the solution that will generate a positive
  semi-definite potential, see eqs.(\ref{eq:solution_BC}).

\item We slightly deform the potential by $B\to B+\delta B$. CMB
  scales start to leave the horizon at $\phi_{\textrm{cmb}}$, which is
  determined by $1+2\eta=n_s$ since still $\epsilon \ll |\eta|$ in all
  cases. Inflation ends at $\phi_{\textrm{end}}$, which is determined
  by $\eta=-1$. We find that both the start point and the end point
  mildly depend on $\delta B$. The correct $\delta B$ is given by
  fixing $N_{\textrm{cmb}}$, for which we consider the range from $45$
  to $65$.
  
\item Having fixed $\delta B$, we can recalculate the slow roll
  parameters at the pivot scale $\eta_{\rm cmb}$ and
  $\epsilon_{\rm cmb}$. We then determine the correct normalization
  $D$ of the potential by requiring $P_{\zeta}=2.1 \times 10^{-9}$.
  This allows us to compute the Hubble value and the inflaton mass.

\end{itemize}

\begin{figure}[h]
\centering
 \includegraphics[width=0.8\textwidth]{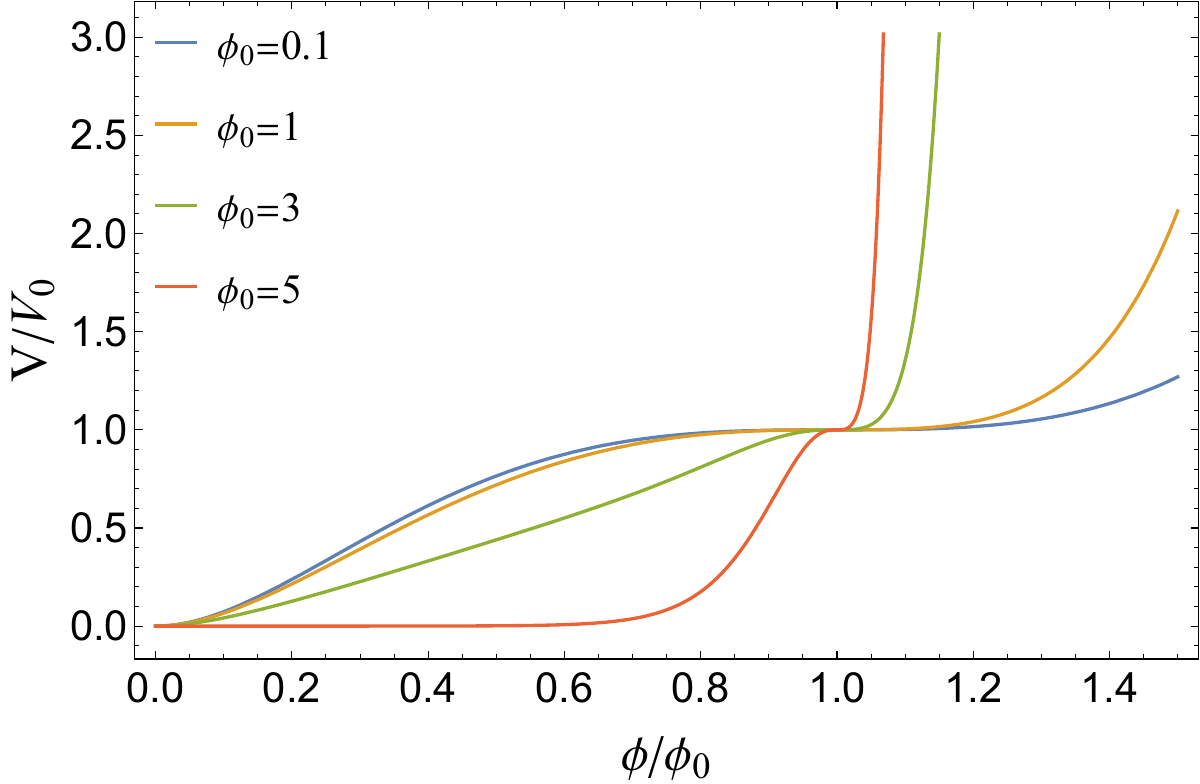}
 \caption{Rescaled inflation potential for different choices of the
   location of the inflection point $\phi_0$. Here $V_0 = V(\phi_0)$
   is the value of the potential at the inflection point. Blue,
   orange, green, and red curves corresponding to $\phi_0=0.1,\,1,\,3$
   and $5$, respectively.}
\label{fig:Vpotential}
\end{figure}

We begin by showing four inflaton potentials with different choices of
$\phi_0$ in Fig.~\ref{fig:Vpotential}. For comparison, these
potentials are rescaled by their values at the inflection point. When
$\phi_0<1$, the shape of the potential becomes independent of $\phi_0$
for $\phi \leq \phi_0$. Increasing $\phi_0$ beyond $1$ shortens the
flat plateau; it also makes it even flatter, which is difficult to see
in this figure.

\begin{figure}[!hbt]
    \centering
\begin{subfigure}{.48\textwidth}
    \centering
    \includegraphics[width=.95\linewidth]{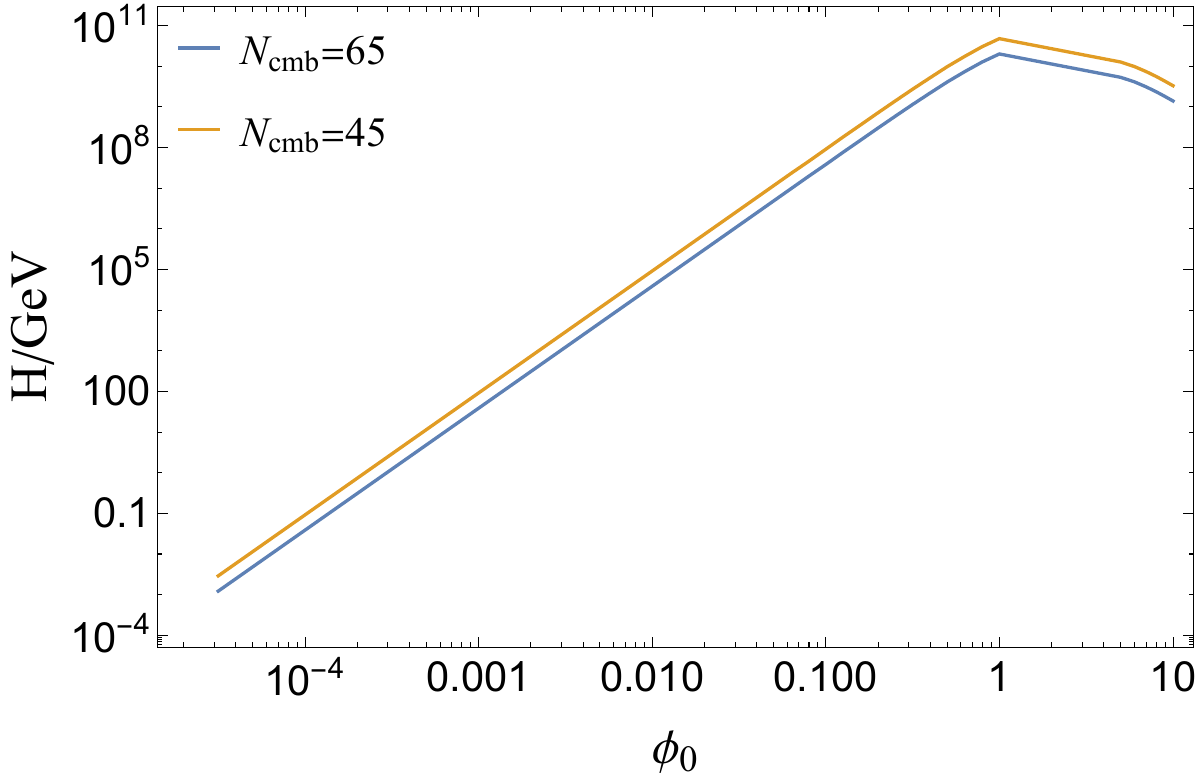}  
    \caption{Hubble scale $H_{\inf}$ during inflation.}
\end{subfigure}
\begin{subfigure}{.48\textwidth}
    \centering
    \includegraphics[width=.95\linewidth]{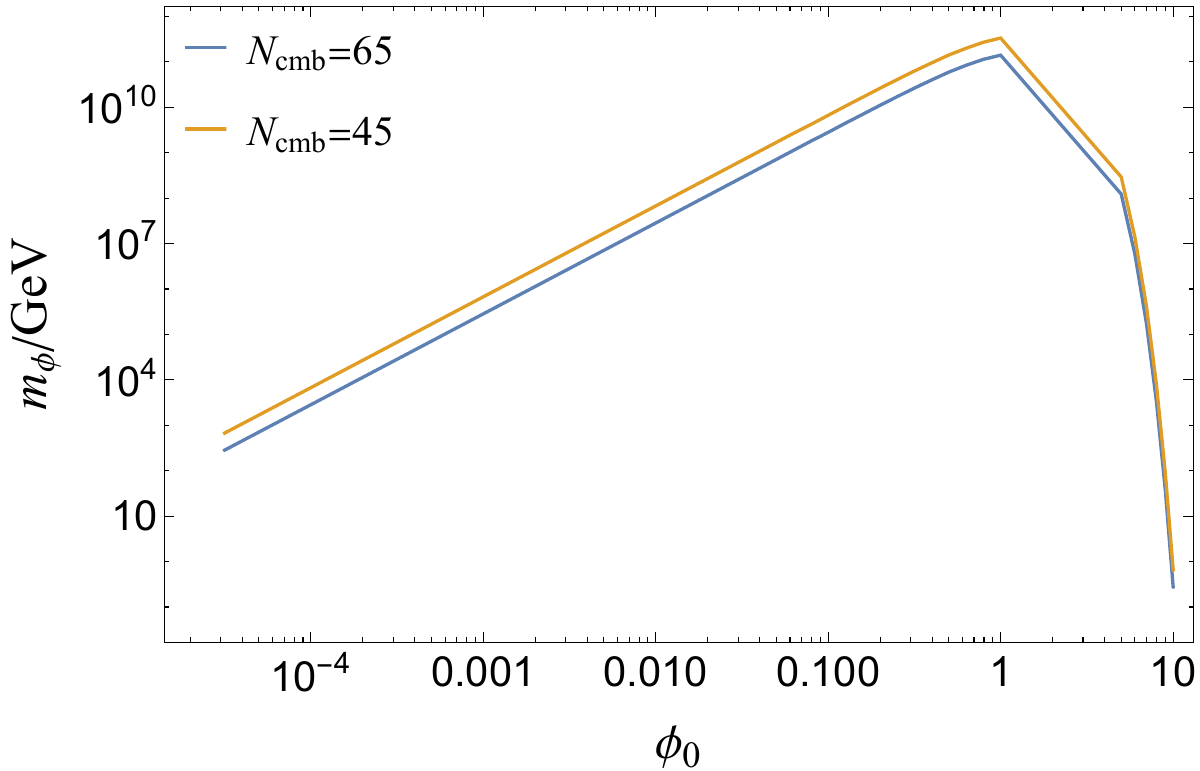}  
    \caption{Inflaton mass $m_\phi$.}
\end{subfigure}
\par\bigskip
\begin{subfigure}{.48\textwidth}
    \centering
    \includegraphics[width=.95\linewidth]{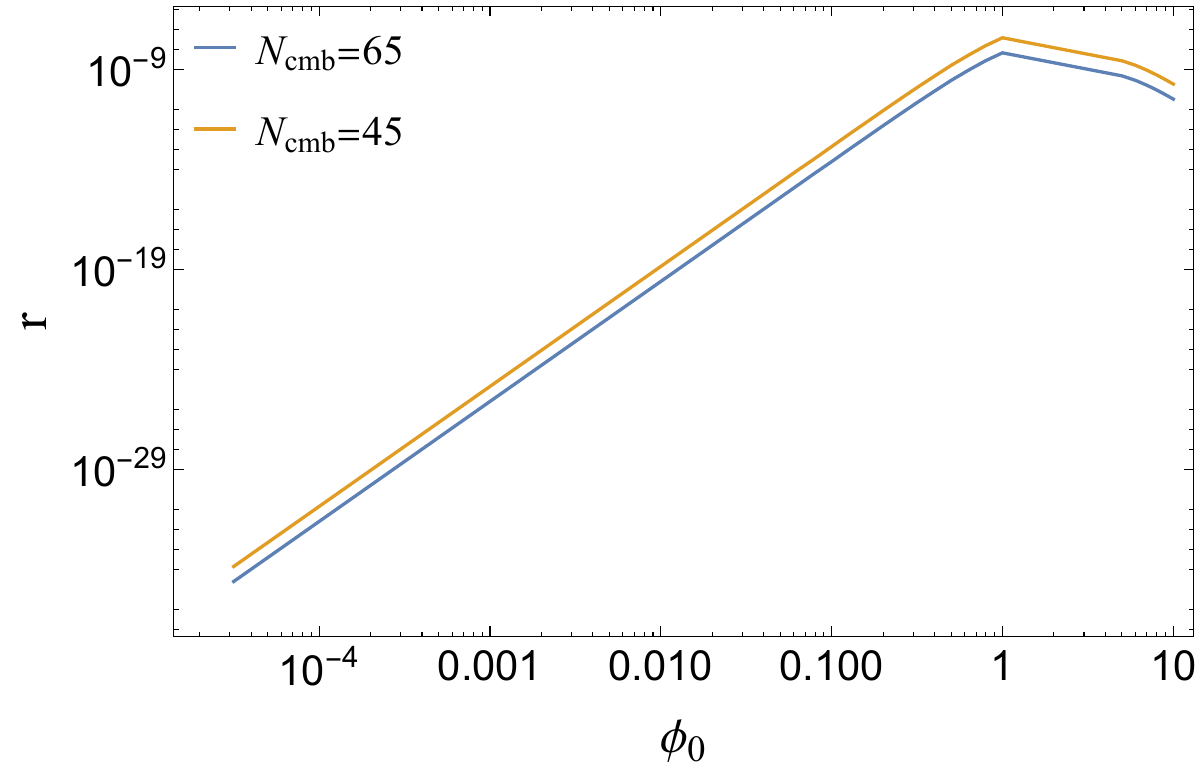}  
    \caption{Tensor to scale ratio $r$.}
\end{subfigure}
\begin{subfigure}{.48\textwidth}
    \centering
    \includegraphics[width=.95\linewidth]{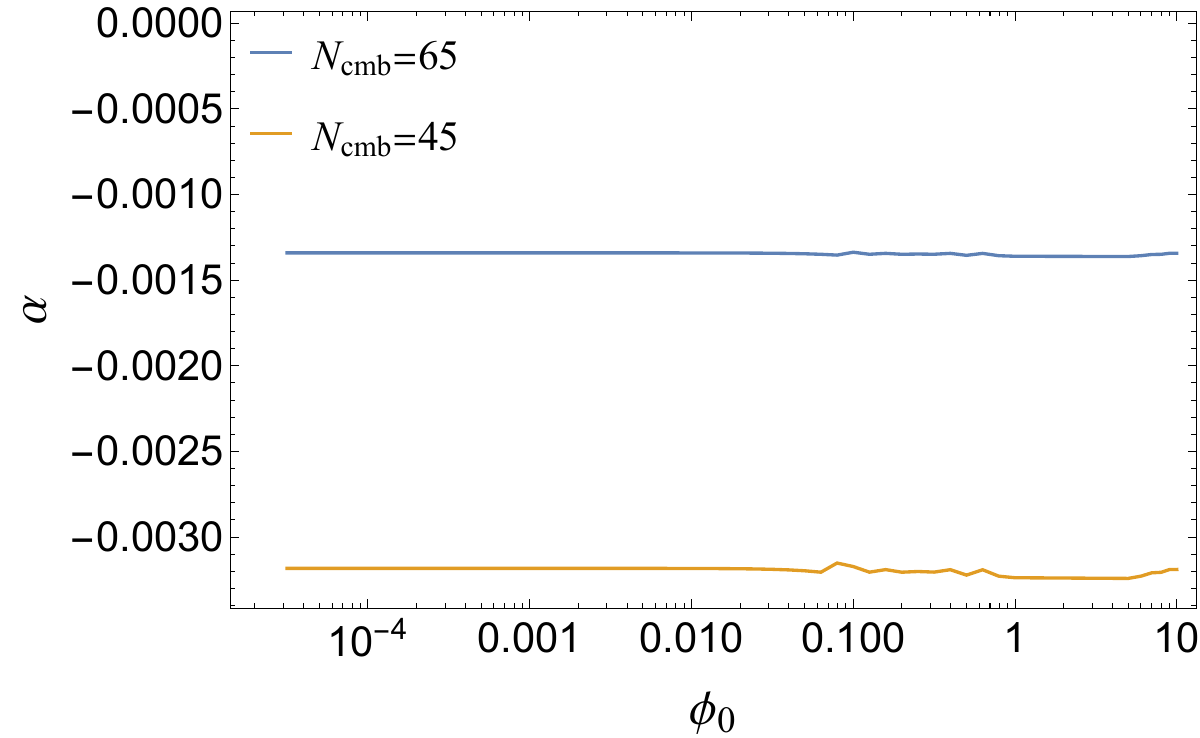}  
    \caption{Running of spectral index $\alpha$.}
\end{subfigure}
\caption{The dependence of the Hubble parameter $H_{\inf}$ during
  inflation, the inflaton mass $m_\phi$, tensor-to-scalar ratio $r$,
  and the running of spectral index $\alpha$ on the position $\phi_0$
  of the inflection point. Different lines represent different choices
  of the number of e-folds: $N_\textrm{cmb}=65$ (blue) and
  $N_\textrm{cmb}=45$ (orange). We fixed $n_s = 0.9659 $ and
  $P_\zeta = 2.1 \times 10^{-9}$.}
\label{fig:sugra_obs}
\end{figure}

The corresponding values of the Hubble parameter during inflation can
be found in the top left frame of Fig.~\ref{fig:sugra_obs}, where we
use blue and orange lines to represent different values of
$N_\textrm{cmb}$. As expected the Hubble scale first increases with
increasing $\phi_0$, then drops once $\phi_0 > 1$. There is no lower
bound on $H_{\rm inf}$ from the pure model perspective. However, the
maximum value is determined by the special shape of the potential and
can never exceed $10^{11}\ \textrm{GeV}$. It obeys a power law when
$\phi_0$ is small, which agrees with our analytic estimation.

The relation between inflaton mass $m_\phi$ and inflection point
$\phi_0$, shown in the top right frame, follows a similar pattern as
the Hubble scale when $\phi_0$ is small, albeit with different
power. However, the inflaton mass drops dramatically for $\phi_0 > 1$,
due to the exponential suppression discussed at the end of the
previous section. When $\phi_0 \approx 10$, the inflaton mass could be
as low as $1\ \textrm{GeV}$. This tiny value differs from the Hubble
scale by more than nine orders of magnitude. Thus our model offers a
way to separate the Hubble and inflaton mass scales.

The running of the spectral index $\alpha$ is shown in the bottom right
frame. It remains independent of $\phi_0$ even for $\phi_0 \geq 1$.
In contrast to the non-SUSY version of the model, $\alpha$ strongly
depends on $N_\textrm{cmb}$. When $N_\textrm{cmb}=45$,
$\alpha \approx -0.0032$. Increasing $N_\textrm{cmb}$ reduces the
absolute value of $\alpha$, reaching $\alpha \approx -0.0013$ for
$N_\textrm{cmb}=65$.

Finally, the bottom left frame of Fig.~\ref{fig:sugra_obs} shows the
relationship between the tensor-to-scalar ratio $r$ and $\phi_0$. It
follows a similar pattern as the Hubble scale. However, since we find
$r < 10^{-7}$, a positive detection of tensor modes by current or
near-future experiments would exclude our model.

In principle one can thus distinguish the small and large field
scenarios through the tensor-to-scalar ratio. If we restrict the
inflection point position so that it should not be much larger than
the Planck scale, $\phi_0<10$, then a small tensor to scalar ratio
$r<10^{-10}$ would favor the small field scenario. If the
tensor-to-scalar ratio is between $10^{-8}$ and $10^{-10}$, we need to
check further the inflation mass. It cannot be measured directly;
however, generally speaking, a large reheating temperature requires a
large inflation mass, which can only be realized in the small field
version of our model. The reheating dynamics can in principle be
probed via the observation of primordial high-frequency gravitational
waves~\cite{Bernal:2023wus,Xu:2024fjl,Jiang:2024akb,Xu:2025wjq}. Unfortunately
none of these observations are feasible with near--future technology.

\subsection{SUSY Breaking by a Polonyi Field}

Clearly SUSY must be broken in any realistic model. In this subsection
we investigate if the existence of a SUSY breaking sector would change
the inflation potential significantly. For simplicity we consider the
classical Polonyi ansatz \cite{Polonyi:1977pj}, where a single chiral
superfield $Z$ with a linear superpotential is introduced to break
SUSY:
\begin{equation} \label{eq:Superpotential}
\begin{split}
  W &= B \Phi^2 + C\Phi^3 + D \Phi^4
  + \mu M_{\textrm{pl}} ( Z + \beta_{\rm P})\\
      &= W_\textrm{I}+W_\textrm{P}\,,
\end{split}
\end{equation}
where $\mu$ essentially sets the SUSY breaking scale; we explicitly
include a factor $M_{\textrm{pl}}$ here to ensure the correct
dimension of $\mu$. Both $\Phi$ and $Z$ are complex fields. As before,
we want the real part of $\Phi$ to be the inflaton field $\phi$. $Z$
is the Polonyi field whose vacuum expectation value
$\langle Z \rangle$ is the only source of SUSY breaking after
inflation, when $\langle \phi \rangle = 0$. SUSY breaking with
vanishing vacuum energy requires $\beta_{\rm P} = 2- \sqrt{3}$, which
gives the gravitino mass $m_{3/2} = \mu e^{2-\sqrt{3}}$ when $Z$ stays
at the SUSY breaking minimum at
$Z = \langle Z \rangle = (\sqrt{3}-1)M_{\textrm{pl}}$.

Let's first consider the case where the Polonyi sector gives a small
perturbation to the inflation potential. For $\phi_0 < 1$ our
previous results suggest:
\begin{equation}
\begin{split}
   \tilde{B} &= \frac{B}{D} \approx 3\phi_0^2 \,,\\
   \tilde{C} &=\frac{C}{D} \approx -2\sqrt{2} \phi_0\,,\\
   D &\approx 4.7 \times 10^{-8}\,.
\end{split}
\end{equation}
We require that the Polonyi field does not change the inflation
potential significantly, which means
$|W_\textrm{I}| \gg |W_\textrm{P}|$ during inflation. The existence of
the SUSY breaking term will not alter the slow roll parameters
significantly if
\begin{equation}
  \epsilon_\mu \ll \epsilon_{\rm cmb} \quad {\rm and} \quad
  \eta_\mu \ll \eta_{\rm cmb}\,,
\end{equation}
where the subscript $\mu$ means the additional contribution to the
slow roll parameter due to the SUSY breaking term.

We assume that during inflation the Polonyi field stays at the origin,
$Z = 0$, which will be verified later. Under this assumption, the
additional contribution to the inflaton potential reads:
\begin{equation}
  V_\mu(\phi) = \frac { {\rm e}^{\frac{1}{2}\phi^2}} {4} \mu \left[
    (-4 \beta_{\rm P} B \phi^2 + 2 \beta_{\rm P} (B+D) \phi^4
    + \sqrt{2} \beta_{\rm P} C \phi^5 + \beta_{\rm P} D \phi^6)
    + \mu ( 4 - 12 \beta_{\rm P}^2 + 2 \beta_{\rm P}^2 \phi^2)\right]\,.
\end{equation}
After substituting $\beta_{\rm P} = 2 - \sqrt{3}$, the additional
contributions to slow roll parameters are:
\begin{equation}\label{eq:SUSYcorrection}
\begin{split}
  \epsilon_\mu \approx &  \sqrt {2 \epsilon_{\rm cmb}} \frac{ \left(
      8 \sqrt{3} - 13\right) \tilde{\mu}^2 \phi_0
    + \left( 4\sqrt{3} - 8\right) \tilde{\mu} \phi_0^3} {2 \phi_0^6}\,,\\
  \eta_\mu \approx & \frac{ (8\sqrt{3} - 13) \tilde{\mu}^2
    + ( 5\sqrt{3} - 10) \tilde{\mu} \phi_0^4} {2\phi_0^6}\,,
\end{split}
\end{equation}
where we have introduced the rescaled parameter $\tilde{\mu}=\mu/D$ to
simplify the expression. The first term in
eq.\eqref{eq:SUSYcorrection} is the cross term between the original
and SUSY breaking induced derivatives of the potential in
$(V')^2$. Using $\epsilon_{\rm cmb} = 8.88 \times 10^{-9} \phi_0^6 $
and requiring $\epsilon_\mu < 0.05 \, \epsilon_\textrm{cmb}$, which
ensures $\eta_\mu \ll \eta_{cmb}$ as well, we get an upper bound on
SUSY breaking scale $\mu$:
\begin{equation} \label{eq:susyestimate}
  \mu < 3.4 \times 10^6\, \left( \frac {\phi_0} {M_{\textrm{pl}}}\right)^6\,
  \textrm{GeV}\,.
\end{equation}
Since we have not found any SUSY particle in collider searches, we
conservatively require $\mu > 1\ \textrm{TeV}$. From
eq.(\ref{eq:susyestimate}) this implies $\phi_0 > 0.3$, corresponding
to $H> 10^8\ \textrm{GeV}.$

\begin{figure}[hbt]
\centering
\includegraphics[width=0.8\textwidth]{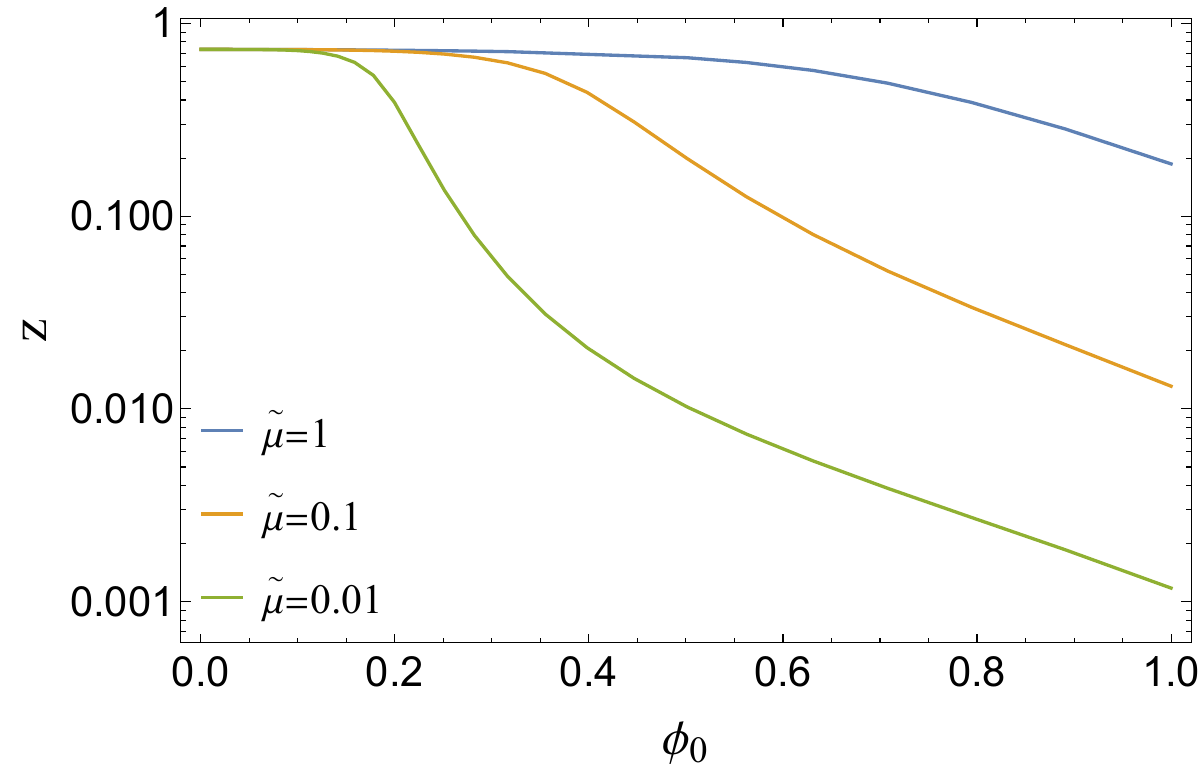}
\caption{Position of the Polonyi field $Z$ during inflation. Different
  colors represent different choices of the relative SUSY breaking
  scale $\tilde{\mu}$. When $\tilde{\mu}\gg \phi_0^2$, the Polonyi
  field stays at $\sqrt{3}-1$, whereas for $\tilde{\mu}\ll \phi_0^2$
  the Polonyi field stays close to the origin.}
\label{fig:position}
\end{figure}

If we increase the SUSY breaking scale while keeping the inflation
scale fixed, the Polonyi field will move from the origin to its
present minimum at $\sqrt{3}-1$. In Fig.~\ref{fig:position} we show
the position of the Polonyi field during inflation, by minimizing
$V(\phi,z)$ with respect to $z$ for fixed $\phi = \phi_0$; here we
only consider the real part $z$ of $Z$, fixing the imaginary part to
the origin throughout. We see that $z$ tends to stay near the origin
when $\tilde{\mu} \ll \phi_0^2$, and slowly moves to its own vacuum at
$\sqrt{3}-1$ as $\tilde{\mu}$ becomes comparable to $\phi_0^2$.

On the other hand, if $|W_\textrm{P}| \gg |W_\textrm{I}|$, the Polonyi
field will stay at $Z=\sqrt{3}-1$; this requires
$|\mu| \gg |D| \phi_0^4$. Focusing on small field solutions, and
imposing the slightly stronger condition $|\mu| \gg |D| \phi_0^2$, the
inflection point conditions $V'(\phi_0) = V''(\phi_0)=0$ up to terms
${\cal O}(\phi_0^2)$ read:
\begin{equation}
\begin{split}
  V'&=0 \Rightarrow 8 \tilde{B}^2 + 18 \sqrt{2} \tilde{B} \tilde{C} \phi_0
  - 4 \left( \sqrt{3} - 2\right) \tilde{B} \tilde{\mu}
  + 18 \tilde{C}^2 \phi_0^2 - 3 \sqrt{2} \left( \sqrt{3} - 3\right)
  \tilde{C} \tilde{\mu} \phi_0 + 2 \tilde{\mu} ^2=0\,,\\
  V''&=0 \Rightarrow 4 \tilde{B}^2 + 18 \sqrt{2} \tilde{B} \tilde{C} \phi_0
  - 2 \left( \sqrt{3} - 2\right ) \tilde{B} \tilde{\mu}
  + 27 \tilde{C}^2 \phi_0 ^2 - 3 \sqrt{2} \left( \sqrt{3} - 3\right)
  \tilde{C} \tilde{\mu}  \phi_0 +\tilde{\mu} ^2=0\,,
\end{split}
\end{equation}
with the following solutions:\footnote{This solution implies
  $B^2 \ll C^2$ as long as $\phi_0 \ll 1$, which a posteriori
  justifies dropping the quartic term $\propto B^2$ in
  eq.(\ref{eq:full_potential}). Moreover, eqs.(\ref{eq:BCtilde}) and
  our assumption $|\tilde \mu| \gg \phi_0^2$ imply that the third
  quartic term in eq.(\ref{eq:full_potential}), $\propto BD$, is also
  negligible. We note that without this additional assumption, higher
  than quartic powers of $\phi$ need to be considered, in which case
  no simple analytical treatment is possible.}
\begin{equation} \label{eq:BCtilde}
  \tilde{B} \approx 0.4952 \tilde{\mu}\,, \quad
  \tilde{C} \approx -\frac{0.4996 \tilde{\mu} }{\phi_0}\,,
\end{equation}
i.e. the ratio $B/C$ again needs to be tuned for inflection point
inflation to occur. The potential at the inflection point is then:
\begin{equation}
  V(\phi_0) \approx 0.1872\, \mu^2 \phi_0^2 \,.
\end{equation}
By Taylor expanding around the inflection point and substituting
$\tilde{B} \to \tilde{B}(1-\delta B), \phi \to \phi_0(1-\delta \phi)$,
we find:
\begin{equation}
 \begin{split}
   \epsilon& \approx \frac{12.1 \left( \delta B + 2.44 \delta \phi^2 \right)^2}
   {\phi_0 ^2}\,,\\
   \eta& \approx \frac{21.6 \delta B - 24.0 \delta \phi}{\phi_0 ^2}\,.
    \end{split}
\end{equation}
This expansion is similar in structure to the previous cases.
Following the same procedure and using
$n_s=0.9659,\, N_{\textrm{cmb}}=65$, we have:
\begin{equation} \label{eq:alphaP}
  \delta B = 4.45 \times 10^{-6} \phi_0^4\,,\quad
  \epsilon_\textrm{cmb} = 3.92\times10^{-10} \phi_0^6, \quad
  \alpha \approx -0.0013\,.
\end{equation}
We can further deduce the scales of SUSY breaking scale and inflation:
\begin{equation}\label{eq:SUSYbreakingresults}
  \mu = \tilde{\mu} D = 4.82 \times 10^{-8} \phi_0^2 \,,  \quad
  H_\textrm{inf} = 4.57\times 10^{-9} \phi_0^3 \,, \quad
  m_\phi \approx 2\mu\,.
\end{equation}
If the Polonyi field already sits in its SUSY breaking minimum during
inflation, all relevant energy scales, i.e., Hubble scale $H_{\inf}$,
the SUSY breaking scale $\mu$, and inflaton mass $m_\phi$, are
completely determined by the position of inflection point
$\phi_0$. The scaling of $H_{\inf}$ and $m_\phi$ with $\phi_0$ is again
as in the non-SUSY version of the model, or as in the SUGRA model
without Polonyi sector. The new feature is that $\phi_0$, or
$H_\textrm{inf}$, also determines $\mu$; again demanding $\mu > 1$ TeV
therefore implies $\phi_0 > 2 \times 10^{-4}$ in this set-up. This
strong correlation can only be relaxed by lifting the Polonyi field
away from the SUSY breaking point $\sqrt{3}-1$.

\begin{figure}[!tbh]
    \centering
\begin{subfigure}{.48\textwidth}
    \centering
    \includegraphics[width=.95\linewidth]{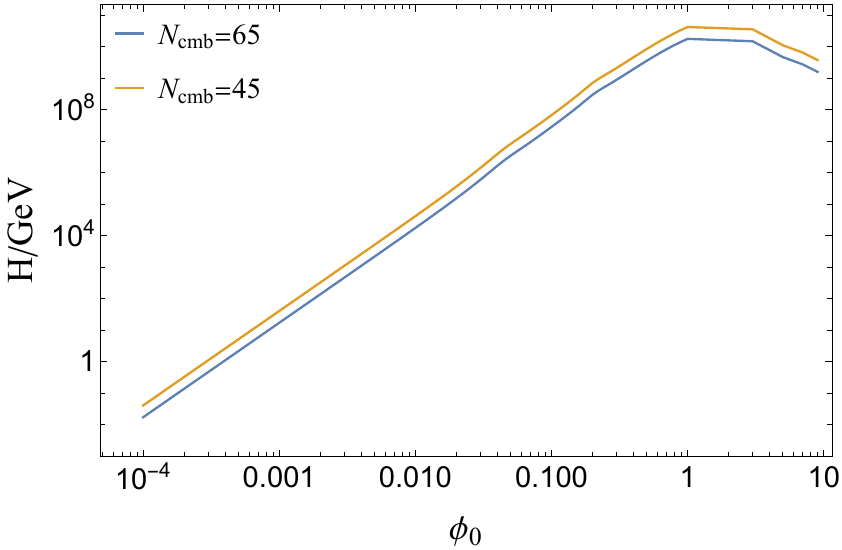}  
    \caption{Hubble scale $H_{\inf}$ during inflation.}
\end{subfigure}
\begin{subfigure}{.48\textwidth}
    \centering
    \includegraphics[width=.95\linewidth]{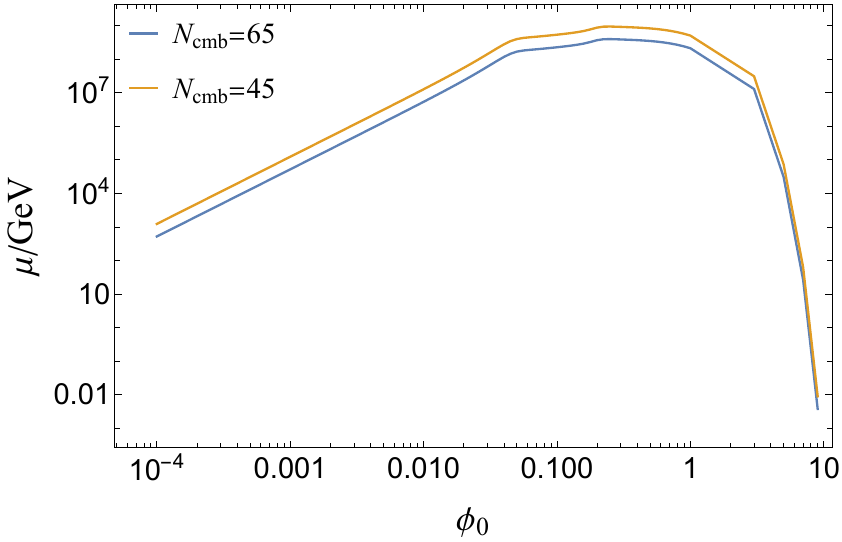}  
    \caption{SUSY breaking scale $\mu$.}
\end{subfigure}
\par\bigskip
\begin{subfigure}{.48\textwidth}
    \centering
    \includegraphics[width=.95\linewidth]{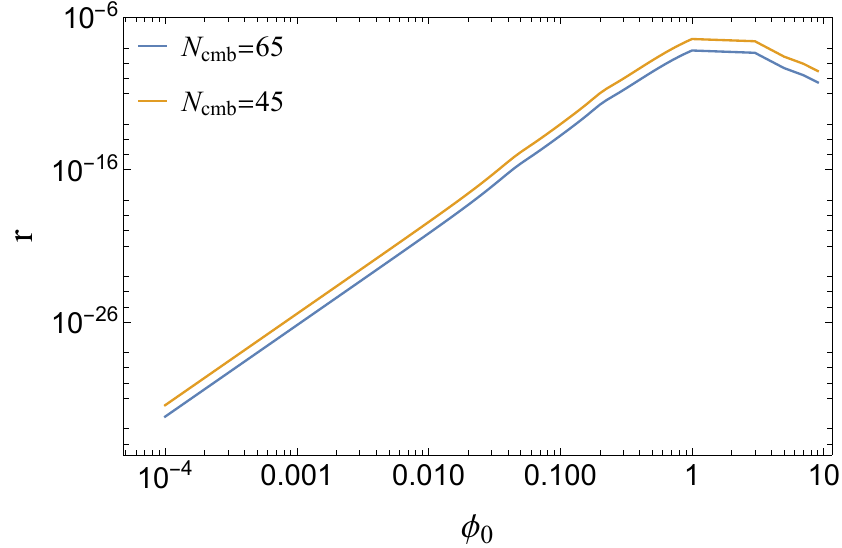}  
    \caption{Tensor to scale ratio $r$.}
\end{subfigure}
\begin{subfigure}{.48\textwidth}
    \centering
    \includegraphics[width=.95\linewidth]{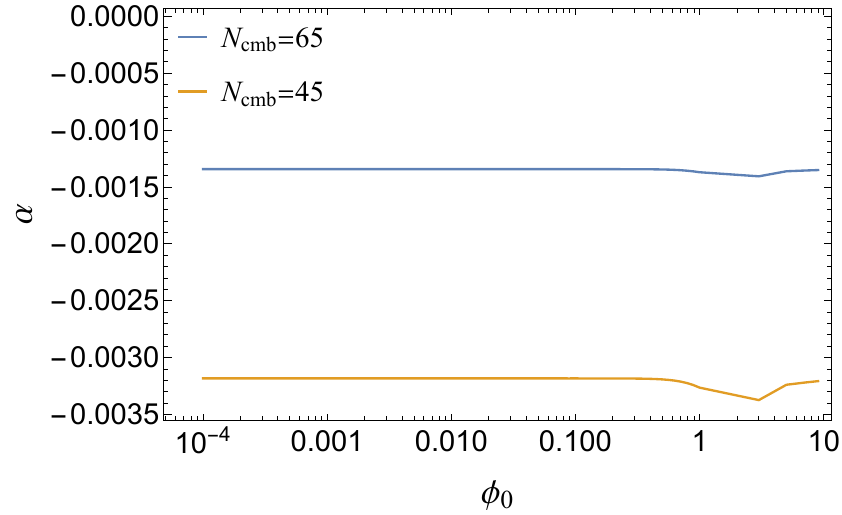}  
    \caption{Running of spectral index $\alpha$.}
\end{subfigure}
\caption{The dependence of the Hubble parameter during inflation
  $H_\textrm{inf}$ (top left), the SUSY breaking scale $\mu$ (top
  right), the tensor-to-scalar ratio $r$ (bottom left), and the
  running of spectral index $\alpha$ (bottom right) on
  $\phi_0$. Different lines represent different choices for the number
  of e-folds: $N_\textrm{cmb}=65$ (blue) and $N_\textrm{cmb}=45$
  (orange). We fixed $\tilde{\mu}=0.01$, $n_s = 0.9659 $ and
  $P_\zeta = 2.1 \times 10^{-9}$ in this graph.}
\label{fig:susybreak_obs}
\end{figure}

In Fig.~\ref{fig:susybreak_obs}, we show how different quantities
depend on $\phi_0$. For $\phi_0 \ll 1$ and $\phi_0^2 \ll \tilde \mu$
this is described by eqs.(\ref{eq:SUSYbreakingresults}) and
(\ref{eq:alphaP}). For our choice $\tilde \mu = 0.01$ the latter
condition is stronger; indeed, we observe some deviations from
eqs.(\ref{eq:SUSYbreakingresults}) once $\phi_0 \gtrsim 0.1$. As
before, we fix the spectral index $n_s$ and the number of e-folds
$N_{\textrm{cmb}}$.

The Hubble parameter $H_\textrm{inf}$ and the tensor-to-scalar ratio
$r$ have the same scaling with $\phi_0$ as before. They both increase
as $\phi_0$ approaches unity, and start to decrease for $\phi_0 >
1$. The running of the spectral index is again almost independent of
$\phi_0$ and of the order of $10^{-3}$. The scaling of the SUSY
breaking scale $\mu$ is rather different. When $\phi_0<0.1$, the
Polonyi field stays around the SUSY breaking point, and increases with
$\phi_0$ as eq.\eqref{eq:SUSYbreakingresults} suggested. When
$\phi_0>0.1$, the Polonyi field is shifted away from the SUSY breaking
point during inflation. This also leads to a milder increase of $\mu$ along
$\phi_0$. Once $\phi_0$ exceeds $1$, the SUSY breaking scale drops
dramatically, which is similar to the behavior of $m_\phi$ in the
previous case. Requiring $\mu \geq 1$ TeV therefore implies $\phi_0
\leq 5$ for this value of $\tilde \mu$.

\begin{figure}[h]
\centering
\includegraphics[width=0.8\textwidth]{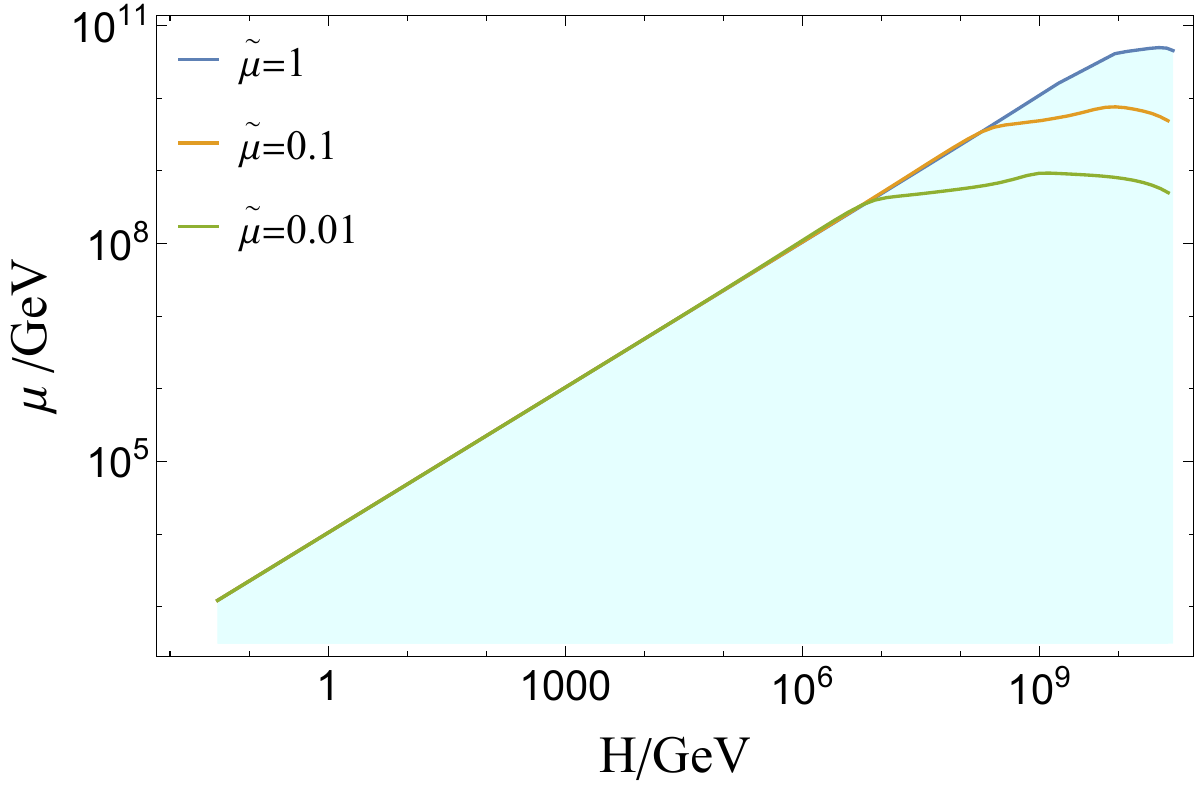}
\caption{The scale of SUSY breaking $\mu$ vs. the inflationary Hubble
  scale $H_{\inf}$ on a log-log scale. Different colors represent different
  choices of relative scale $\tilde{\mu}$. The straight line is the
  Polonyi field dominated case, where the SUSY breaking scale only
  depends on the inflection point positions. The right, flat region is
  an inflaton field dominated region, where the SUSY breaking scale
  depends linearly on the relative scale $\tilde{\mu}$.}
\label{fig:susyh}
\end{figure}

So far we fixed $\tilde \mu = 0.01$. For $\phi_0^2 \ll \tilde \mu$
this choice is in fact irrelevant, since the physical parameters
$B, \, \delta B, \, C$ and $\mu$ are all fixed uniquely for given
$\phi_0$, see eqs.(\ref{eq:BCtilde}), (\ref{eq:alphaP}) and
(\ref{eq:SUSYbreakingresults}). However, we see in
Fig.~\ref{fig:susyh}, which shows the relation between $\mu$ and
$H_{\inf}$, that this is no longer true for
$\phi_0 \geq \sqrt{\tilde\mu}$. Nevertheless, this figure shows that
for a given Hubble scale during inflation there will be a maximum SUSY
breaking scale it can host, given by the analytical solution of
eq.(\ref{eq:SUSYbreakingresults}):
\begin{equation}
  \mu < \frac {10.5} {\phi_0} H_{\inf} \approx 2.3 \times 10^4 \left(
    \frac{H_{\inf}}{\textrm{GeV}} \right)^{2/3} \textrm{GeV}\,.
\end{equation}
Recall that this corresponds to a scenario where the Polonyi field
remains at its current value throughout inflation.  If it moves away
from this value during inflation, one will have more freedom to set
the value of $\mu$, depending on the position of the Polonyi field
during inflation. In the opposite extreme, where the Polonyi field
stays at the origin during inflation and only perturbs the inflaton
potential, the SUSY breaking scale would simply be
$\mu=\tilde{\mu}\times D$, where $D$ can be treated as a
constant. This explains why the relative ratio of three different
cases when they deviate from the straight line in Fig.~\ref{fig:susyh}
is almost a constant.

We conclude that only the light cyan region below the topmost line in
Fig.~\ref{fig:susyh} is accessible in our model. Different choices of
$\tilde{\mu}$ will leave the straight line at different Hubble scales,
and can thus populate this region, always keeping in mind that
$\mu > 1$ TeV is needed for phenomenological reasons. The resulting
lower bound on the Hubble scale during inflation is around
$1~\textrm{GeV}$. This bound is much smaller than the naive estimate
of $10^8$ GeV we derived below eq.\eqref{eq:susyestimate} from the
requirement that the Polonyi sector can be treated as a small
perturbation. 

\section{Summary}

In this paper we revisited the inflection point
inflation model in the SUGRA framework. We adopted the minimal
assumption that only one canonical field drives inflation. While
supersymmetry protects the flatness of the potential from radiative
corrections, local SUSY or SUGRA also modifies the potential through
non-renormalizable terms. These new terms contribute to slow roll
parameters on an equal footing. As in the non-supersymmetric case the
shape of the potential is determined by the position $\phi_0$ of the
inflection point, which is a free parameter of our model. When fixing
the well-constrained power spectrum of curvature perturbations and its
spectral index, we find $\phi_0$ controls the tensor-to-scalar ratio
$r$, the Hubble scale during inflation $H_\textrm{inf}$, and the
physical inflaton mass $m_\phi$.

For $\phi_0 \ll 1$ a perturbative treatment is possible. In this case,
$r,\, H_\textrm{inf}$ and $m_\phi$ are monomial functions of $\phi_0$
and reach their maximum around $\phi_0 \approx 1$. The running of the
spectral index $\alpha$ is almost independent of $\phi_0$ but depends
more strongly on $N_\textrm{cmb}$ than in the non-supersymmetric version of the model. The tensor-to-scalar ratio $r$ is
always smaller than $10^{-7}$, which is below the sensitivity of any
current or planned experiments. The running of the spectral index,
which is $\mathcal{O}(-10^{-3})$, might be probed by the next
generation of CMB experiments. The predictions of this SUGRA model are
quite similar to those from the non-supersymmetric model. All observables
have the same scaling with respect to $\phi_0$. Thus, even though the
SUGRA potential contains terms up to $\phi^6$ while the non-supersymmetric potential only has terms up to $\phi^4$, the latter still provides a relatively reliable estimate of inflationary quantities.

The main difference between the SUGRA case and the non-supersymmetric case
appears when $\phi_0$ exceeds $1$. In this region the exponential
factor $e^{\phi^2/2}$ in the SUGRA case becomes large, which
suppresses $r,\, H_\textrm{inf}$ and $m_\phi$. The energy scales are
therefore bounded from above: $H_\textrm{inf}<10^{11}\ \textrm{GeV}$
and $m_\phi<10^{12}\ \textrm{GeV}$. The non-supersymmetric potential is
not able to capture this behavior, which leads to a very different
prediction of inflationary observables in this large field scenario.

We further added a SUSY breaking Polonyi sector to the model. If the
SUSY breaking scale is much smaller than the Hubble scale, the Polonyi
field will stay at the origin and serve as a perturbation to the
field. When these two energy scales become comparable, the Polonyi
field will move away from the origin and modify the inflation
potential. In the opposite extreme the Polonyi field stays close to
its post-inflationary value throughout. These effects lead to a
nontrivial bound between the SUSY breaking scale and the inflation
scale. We find that for a TeV scale SUSY breaking, we need the Hubble
scale to be larger than $1$ GeV.

It has been pointed out that in the KKLT model, the Hubble scale is
always smaller than the gravitino mass $m_{3/2}$ (or SUSY breaking
scale $\mu$)~\cite{Kallosh:2004yh}. In our model, we find a slightly
different conclusion: in some regions, the Hubble scale can be larger
than the gravitino mass. In such a scenario supersymmetry will protect
the inflaton superpotential from loop corrections. This should allow
larger couplings of the inflaton to Standard Model (super)fields, and
thus larger reheat temperature than in the non--supersymmetric
version.

\bibliographystyle{JHEP}
\bibliography{ref}

\providecommand{\href}[2]{#2}\begingroup\raggedright\begin{thebibliography}{10}

\bibitem{Planck:2018jri}
{\bf Planck} Collaboration, Y.~Akrami et~al., {\it {Planck 2018 results. X. Constraints on inflation}},  {\em Astron. Astrophys.} {\bf 641} (2020) A10, [\href{http://arxiv.org/abs/1807.06211}{{\tt arXiv:1807.06211}}].

\bibitem{BICEP:2021xfz}
{\bf BICEP, Keck} Collaboration, P.~A.~R. Ade et~al., {\it {Improved Constraints on Primordial Gravitational Waves using Planck, WMAP, and BICEP/Keck Observations through the 2018 Observing Season}},  {\em Phys. Rev. Lett.} {\bf 127} (2021), no.~15 151301, [\href{http://arxiv.org/abs/2110.00483}{{\tt arXiv:2110.00483}}].

\bibitem{Starobinsky:1980te}
A.~A. Starobinsky, {\it {A New Type of Isotropic Cosmological Models Without Singularity}},  {\em Phys. Lett. B} {\bf 91} (1980) 99--102.

\bibitem{Guth:1980zm}
A.~H. Guth, {\it {The Inflationary Universe: A Possible Solution to the Horizon and Flatness Problems}},  {\em Phys. Rev. D} {\bf 23} (1981) 347--356.

\bibitem{Linde:1981mu}
A.~D. Linde, {\it {A New Inflationary Universe Scenario: A Possible Solution of the Horizon, Flatness, Homogeneity, Isotropy and Primordial Monopole Problems}},  {\em Phys. Lett. B} {\bf 108} (1982) 389--393.

\bibitem{Albrecht:1982wi}
A.~Albrecht and P.~J. Steinhardt, {\it {Cosmology for Grand Unified Theories with Radiatively Induced Symmetry Breaking}},  {\em Phys. Rev. Lett.} {\bf 48} (1982) 1220--1223.

\bibitem{Lyth:2009imm}
D.~H. Lyth and A.~R. Liddle, {\em {The Primordial Density Perturbation}}.
\newblock Cambridge Univ. Press, Cambridge, UK, 2009.

\bibitem{ACT:2025fju}
{\bf ACT} Collaboration, T.~Louis et~al., {\it {The Atacama Cosmology Telescope: DR6 Power Spectra, Likelihoods and $\Lambda$CDM Parameters}},  \href{http://arxiv.org/abs/2503.14452}{{\tt arXiv:2503.14452}}.

\bibitem{ACT:2025tim}
{\bf ACT} Collaboration, E.~Calabrese et~al., {\it {The Atacama Cosmology Telescope: DR6 Constraints on Extended Cosmological Models}},  \href{http://arxiv.org/abs/2503.14454}{{\tt arXiv:2503.14454}}.

\bibitem{Allahverdi:2006iq}
R.~Allahverdi, K.~Enqvist, J.~Garcia-Bellido, and A.~Mazumdar, {\it {Gauge invariant MSSM inflaton}},  {\em Phys. Rev. Lett.} {\bf 97} (2006) 191304, [\href{http://arxiv.org/abs/hep-ph/0605035}{{\tt hep-ph/0605035}}].

\bibitem{Allahverdi:2006cx}
R.~Allahverdi, A.~Kusenko, and A.~Mazumdar, {\it {A-term inflation and the smallness of neutrino masses}},  {\em JCAP} {\bf 07} (2007) 018, [\href{http://arxiv.org/abs/hep-ph/0608138}{{\tt hep-ph/0608138}}].

\bibitem{Allahverdi:2006we}
R.~Allahverdi, K.~Enqvist, J.~Garcia-Bellido, A.~Jokinen, and A.~Mazumdar, {\it {MSSM flat direction inflation: Slow roll, stability, fine tunning and reheating}},  {\em JCAP} {\bf 06} (2007) 019, [\href{http://arxiv.org/abs/hep-ph/0610134}{{\tt hep-ph/0610134}}].

\bibitem{Enqvist:2010vd}
K.~Enqvist, A.~Mazumdar, and P.~Stephens, {\it {Inflection point inflation within supersymmetry}},  {\em JCAP} {\bf 06} (2010) 020, [\href{http://arxiv.org/abs/1004.3724}{{\tt arXiv:1004.3724}}].

\bibitem{Okada:2016ssd}
N.~Okada and D.~Raut, {\it {Inflection-point Higgs Inflation}},  {\em Phys. Rev. D} {\bf 95} (2017), no.~3 035035, [\href{http://arxiv.org/abs/1610.09362}{{\tt arXiv:1610.09362}}].

\bibitem{Okada:2017cvy}
N.~Okada, S.~Okada, and D.~Raut, {\it {Inflection-point inflation in hyper-charge oriented U(1)$_X$ model}},  {\em Phys. Rev. D} {\bf 95} (2017), no.~5 055030, [\href{http://arxiv.org/abs/1702.02938}{{\tt arXiv:1702.02938}}].

\bibitem{Baumann:2007np}
D.~Baumann, A.~Dymarsky, I.~R. Klebanov, L.~McAllister, and P.~J. Steinhardt, {\it {A Delicate universe}},  {\em Phys. Rev. Lett.} {\bf 99} (2007) 141601, [\href{http://arxiv.org/abs/0705.3837}{{\tt arXiv:0705.3837}}].

\bibitem{Baumann:2007ah}
D.~Baumann, A.~Dymarsky, I.~R. Klebanov, and L.~McAllister, {\it {Towards an Explicit Model of D-brane Inflation}},  {\em JCAP} {\bf 01} (2008) 024, [\href{http://arxiv.org/abs/0706.0360}{{\tt arXiv:0706.0360}}].

\bibitem{Hotchkiss:2011am}
S.~Hotchkiss, A.~Mazumdar, and S.~Nadathur, {\it {Inflection point inflation: WMAP constraints and a solution to the fine-tuning problem}},  {\em JCAP} {\bf 06} (2011) 002, [\href{http://arxiv.org/abs/1101.6046}{{\tt arXiv:1101.6046}}].

\bibitem{Chang:2013cba}
H.-Y. Chang and R.~J. Scherrer, {\it {Inflection Point Quintessence}},  {\em Phys. Rev. D} {\bf 88} (2013), no.~8 083003, [\href{http://arxiv.org/abs/1306.4662}{{\tt arXiv:1306.4662}}].

\bibitem{Martin:2013tda}
J.~Martin, C.~Ringeval, and V.~Vennin, {\it {Encyclop\ae{}dia Inflationaris}: {Opiparous Edition}},  {\em Phys. Dark Univ.} {\bf 5-6} (2014) 75--235, [\href{http://arxiv.org/abs/1303.3787}{{\tt arXiv:1303.3787}}].

\bibitem{Martin:2013nzq}
J.~Martin, C.~Ringeval, R.~Trotta, and V.~Vennin, {\it {The Best Inflationary Models After Planck}},  {\em JCAP} {\bf 03} (2014) 039, [\href{http://arxiv.org/abs/1312.3529}{{\tt arXiv:1312.3529}}].

\bibitem{Choi:2016eif}
S.-M. Choi and H.~M. Lee, {\it {Inflection point inflation and reheating}},  {\em Eur. Phys. J. C} {\bf 76} (2016), no.~6 303, [\href{http://arxiv.org/abs/1601.05979}{{\tt arXiv:1601.05979}}].

\bibitem{Dimopoulos:2017xox}
K.~Dimopoulos, C.~Owen, and A.~Racioppi, {\it {Loop inflection-point inflation}},  {\em Astropart. Phys.} {\bf 103} (2018) 16--20, [\href{http://arxiv.org/abs/1706.09735}{{\tt arXiv:1706.09735}}].

\bibitem{Musoke:2017frr}
N.~Musoke and R.~Easther, {\it {Expectations for Inflationary Observables: Simple or Natural?}},  {\em JCAP} {\bf 12} (2017) 032, [\href{http://arxiv.org/abs/1709.01192}{{\tt arXiv:1709.01192}}].

\bibitem{Drees:2021wgd}
M.~Drees and Y.~Xu, {\it {Small field polynomial inflation: reheating, radiative stability and lower bound}},  {\em JCAP} {\bf 09} (2021) 012, [\href{http://arxiv.org/abs/2104.03977}{{\tt arXiv:2104.03977}}].

\bibitem{Drees:2022aea}
M.~Drees and Y.~Xu, {\it {Large field polynomial inflation: parameter space, predictions and (double) eternal nature}},  {\em JCAP} {\bf 12} (2022) 005, [\href{http://arxiv.org/abs/2209.07545}{{\tt arXiv:2209.07545}}].

\bibitem{Drees:2024hok}
M.~Drees and Y.~Xu, {\it {Parameter space of leptogenesis in polynomial inflation}},  {\em JCAP} {\bf 04} (2024) 036, [\href{http://arxiv.org/abs/2401.02485}{{\tt arXiv:2401.02485}}].

\bibitem{Bernal:2024ykj}
N.~Bernal, J.~Harz, M.~A. Mojahed, and Y.~Xu, {\it {Graviton- and inflaton-mediated dark matter production after large field polynomial inflation}},  {\em Phys. Rev. D} {\bf 111} (2025), no.~4 043517, [\href{http://arxiv.org/abs/2406.19447}{{\tt arXiv:2406.19447}}].

\bibitem{Drees:2025iue}
M.~Drees and C.~Wang, {\it {Inflaton Self Resonance, Oscillons, and Gravitational Waves in Small Field Polynomial Inflation}},  \href{http://arxiv.org/abs/2501.13811}{{\tt arXiv:2501.13811}}.

\bibitem{Nanopoulos:1982bv}
D.~V. Nanopoulos, K.~A. Olive, M.~Srednicki, and K.~Tamvakis, {\it {Primordial Inflation in Simple Supergravity}},  {\em Phys. Lett. B} {\bf 123} (1983) 41--44.

\bibitem{Adams:1992bn}
F.~C. Adams, J.~R. Bond, K.~Freese, J.~A. Frieman, and A.~V. Olinto, {\it {Natural inflation: Particle physics models, power law spectra for large scale structure, and constraints from COBE}},  {\em Phys. Rev. D} {\bf 47} (1993) 426--455, [\href{http://arxiv.org/abs/hep-ph/9207245}{{\tt hep-ph/9207245}}].

\bibitem{Copeland:1994vg}
E.~J. Copeland, A.~R. Liddle, D.~H. Lyth, E.~D. Stewart, and D.~Wands, {\it {False vacuum inflation with Einstein gravity}},  {\em Phys. Rev. D} {\bf 49} (1994) 6410--6433, [\href{http://arxiv.org/abs/astro-ph/9401011}{{\tt astro-ph/9401011}}].

\bibitem{Kumekawa:1994gx}
K.~Kumekawa, T.~Moroi, and T.~Yanagida, {\it {Flat potential for inflaton with a discrete R invariance in supergravity}},  {\em Prog. Theor. Phys.} {\bf 92} (1994) 437--448, [\href{http://arxiv.org/abs/hep-ph/9405337}{{\tt hep-ph/9405337}}].

\bibitem{Izawa:1996dv}
K.~I. Izawa and T.~Yanagida, {\it {Natural new inflation in broken supergravity}},  {\em Phys. Lett. B} {\bf 393} (1997) 331--336, [\href{http://arxiv.org/abs/hep-ph/9608359}{{\tt hep-ph/9608359}}].

\bibitem{Panagiotakopoulos:1997qd}
C.~Panagiotakopoulos, {\it {Blue perturbation spectra from hybrid inflation with canonical supergravity}},  {\em Phys. Rev. D} {\bf 55} (1997) R7335--R7339, [\href{http://arxiv.org/abs/hep-ph/9702433}{{\tt hep-ph/9702433}}].

\bibitem{Linde:1997sj}
A.~D. Linde and A.~Riotto, {\it {Hybrid inflation in supergravity}},  {\em Phys. Rev. D} {\bf 56} (1997) R1841--R1844, [\href{http://arxiv.org/abs/hep-ph/9703209}{{\tt hep-ph/9703209}}].

\bibitem{Kawasaki:2000yn}
M.~Kawasaki, M.~Yamaguchi, and T.~Yanagida, {\it {Natural chaotic inflation in supergravity}},  {\em Phys. Rev. Lett.} {\bf 85} (2000) 3572--3575, [\href{http://arxiv.org/abs/hep-ph/0004243}{{\tt hep-ph/0004243}}].

\bibitem{Binetruy:2004hh}
P.~Binetruy, G.~Dvali, R.~Kallosh, and A.~Van~Proeyen, {\it {Fayet-Iliopoulos terms in supergravity and cosmology}},  {\em Class. Quant. Grav.} {\bf 21} (2004) 3137--3170, [\href{http://arxiv.org/abs/hep-th/0402046}{{\tt hep-th/0402046}}].

\bibitem{Ellis:2013xoa}
J.~Ellis, D.~V. Nanopoulos, and K.~A. Olive, {\it {No-Scale Supergravity Realization of the Starobinsky Model of Inflation}},  {\em Phys. Rev. Lett.} {\bf 111} (2013) 111301, [\href{http://arxiv.org/abs/1305.1247}{{\tt arXiv:1305.1247}}]. [Erratum: Phys.Rev.Lett. 111, 129902 (2013)].

\bibitem{Ellis:2013nxa}
J.~Ellis, D.~V. Nanopoulos, and K.~A. Olive, {\it {Starobinsky-like Inflationary Models as Avatars of No-Scale Supergravity}},  {\em JCAP} {\bf 10} (2013) 009, [\href{http://arxiv.org/abs/1307.3537}{{\tt arXiv:1307.3537}}].

\bibitem{Kallosh:2013yoa}
R.~Kallosh, A.~Linde, and D.~Roest, {\it {Superconformal Inflationary $\alpha$-Attractors}},  {\em JHEP} {\bf 11} (2013) 198, [\href{http://arxiv.org/abs/1311.0472}{{\tt arXiv:1311.0472}}].

\bibitem{Carrasco:2015pla}
J.~J.~M. Carrasco, R.~Kallosh, and A.~Linde, {\it {$\alpha $-Attractors: Planck, LHC and Dark Energy}},  {\em JHEP} {\bf 10} (2015) 147, [\href{http://arxiv.org/abs/1506.01708}{{\tt arXiv:1506.01708}}].

\bibitem{Holman:1984yj}
R.~Holman, P.~Ramond, and G.~G. Ross, {\it {Supersymmetric Inflationary Cosmology}},  {\em Phys. Lett. B} {\bf 137} (1984) 343--347.

\bibitem{Ross:1995dq}
G.~G. Ross and S.~Sarkar, {\it {Successful supersymmetric inflation}},  {\em Nucl. Phys. B} {\bf 461} (1996) 597--624, [\href{http://arxiv.org/abs/hep-ph/9506283}{{\tt hep-ph/9506283}}].

\bibitem{Mazumdar:2011ih}
A.~Mazumdar, S.~Nadathur, and P.~Stephens, {\it {Inflation with large supergravity corrections}},  {\em Phys. Rev. D} {\bf 85} (2012) 045001, [\href{http://arxiv.org/abs/1105.0430}{{\tt arXiv:1105.0430}}].

\bibitem{Linde:2007jn}
A.~D. Linde and A.~Westphal, {\it {Accidental Inflation in String Theory}},  {\em JCAP} {\bf 03} (2008) 005, [\href{http://arxiv.org/abs/0712.1610}{{\tt arXiv:0712.1610}}].

\bibitem{Conlon:2008cj}
J.~P. Conlon, R.~Kallosh, A.~D. Linde, and F.~Quevedo, {\it {Volume Modulus Inflation and the Gravitino Mass Problem}},  {\em JCAP} {\bf 09} (2008) 011, [\href{http://arxiv.org/abs/0806.0809}{{\tt arXiv:0806.0809}}].

\bibitem{Maharana:2015saa}
A.~Maharana, M.~Rummel, and Y.~Sumitomo, {\it {Accidental K{\"a}hler moduli inflation}},  {\em JCAP} {\bf 09} (2015) 040, [\href{http://arxiv.org/abs/1504.07202}{{\tt arXiv:1504.07202}}].

\bibitem{Gao:2015yha}
T.-J. Gao and Z.-K. Guo, {\it {Inflection point inflation and dark energy in supergravity}},  {\em Phys. Rev. D} {\bf 91} (2015) 123502, [\href{http://arxiv.org/abs/1503.05643}{{\tt arXiv:1503.05643}}].

\bibitem{Pallis:2023aom}
C.~Pallis, {\it {Inflection-point sgoldstino inflation in no-scale supergravity}},  {\em Phys. Lett. B} {\bf 843} (2023) 138018, [\href{http://arxiv.org/abs/2302.12214}{{\tt arXiv:2302.12214}}].

\bibitem{Gao:2016xfv}
T.-J. Gao, W.-T. Xu, and X.-Y. Yang, {\it {Inflection point in running kinetic term inflation}},  {\em Mod. Phys. Lett. A} {\bf 32} (2017), no.~2 1750072, [\href{http://arxiv.org/abs/1606.05951}{{\tt arXiv:1606.05951}}].

\bibitem{Aldabergenov:2020bpt}
Y.~Aldabergenov, A.~Addazi, and S.~V. Ketov, {\it {Primordial black holes from modified supergravity}},  {\em Eur. Phys. J. C} {\bf 80} (2020), no.~10 917, [\href{http://arxiv.org/abs/2006.16641}{{\tt arXiv:2006.16641}}].

\bibitem{Badziak:2008yg}
M.~Badziak and M.~Olechowski, {\it {Volume modulus inflation and a low scale of SUSY breaking}},  {\em JCAP} {\bf 07} (2008) 021, [\href{http://arxiv.org/abs/0802.1014}{{\tt arXiv:0802.1014}}].

\bibitem{Ben-Dayan:2008fhy}
I.~Ben-Dayan, R.~Brustein, and S.~P. de~Alwis, {\it {Models of Modular Inflation and Their Phenomenological Consequences}},  {\em JCAP} {\bf 07} (2008) 011, [\href{http://arxiv.org/abs/0802.3160}{{\tt arXiv:0802.3160}}].

\bibitem{Covi:2008cn}
L.~Covi, M.~Gomez-Reino, C.~Gross, J.~Louis, G.~A. Palma, and C.~A. Scrucca, {\it {Constraints on modular inflation in supergravity and string theory}},  {\em JHEP} {\bf 08} (2008) 055, [\href{http://arxiv.org/abs/0805.3290}{{\tt arXiv:0805.3290}}].

\bibitem{Ketov:2014qha}
S.~V. Ketov and T.~Terada, {\it {Inflation in supergravity with a single chiral superfield}},  {\em Phys. Lett. B} {\bf 736} (2014) 272--277, [\href{http://arxiv.org/abs/1406.0252}{{\tt arXiv:1406.0252}}].

\bibitem{Polonyi:1977pj}
J.~Polonyi, {\it {Generalization of the Massive Scalar Multiplet Coupling to the Supergravity}}, .

\bibitem{Ellis:1986zt}
J.~R. Ellis, D.~V. Nanopoulos, and M.~Quiros, {\it {On the Axion, Dilaton, Polonyi, Gravitino and Shadow Matter Problems in Supergravity and Superstring Models}},  {\em Phys. Lett. B} {\bf 174} (1986) 176--182.

\bibitem{Coughlan:1983ci}
G.~D. Coughlan, W.~Fischler, E.~W. Kolb, S.~Raby, and G.~G. Ross, {\it {Cosmological Problems for the Polonyi Potential}},  {\em Phys. Lett. B} {\bf 131} (1983) 59--64.

\bibitem{Ibe:2006am}
M.~Ibe, Y.~Shinbara, and T.~T. Yanagida, {\it {The Polonyi Problem and Upper bound on Inflation Scale in Supergravity}},  {\em Phys. Lett. B} {\bf 639} (2006) 534--540, [\href{http://arxiv.org/abs/hep-ph/0605252}{{\tt hep-ph/0605252}}].

\bibitem{Addazi:2017ulg}
A.~Addazi, S.~V. Ketov, and M.~Y. Khlopov, {\it {Gravitino and Polonyi production in supergravity}},  {\em Eur. Phys. J. C} {\bf 78} (2018), no.~8 642, [\href{http://arxiv.org/abs/1708.05393}{{\tt arXiv:1708.05393}}].

\bibitem{Aldabergenov:2017bjt}
Y.~Aldabergenov and S.~V. Ketov, {\it {Higgs mechanism and cosmological constant in $N=1$ supergravity with inflaton in a vector multiplet}},  {\em Eur. Phys. J. C} {\bf 77} (2017), no.~4 233, [\href{http://arxiv.org/abs/1701.08240}{{\tt arXiv:1701.08240}}].

\bibitem{Aldabergenov:2017hvp}
Y.~Aldabergenov and S.~V. Ketov, {\it {Removing instability of inflation in Polonyi{\textendash}Starobinsky supergravity by adding FI term}},  {\em Mod. Phys. Lett. A} {\bf 33} (2018), no.~5 1850032, [\href{http://arxiv.org/abs/1711.06789}{{\tt arXiv:1711.06789}}].

\bibitem{Abe:2018rnu}
H.~Abe, Y.~Aldabergenov, S.~Aoki, and S.~V. Ketov, {\it {Polonyi{\textendash}Starobinsky supergravity with inflaton in a massive vector multiplet with DBI and FI terms}},  {\em Class. Quant. Grav.} {\bf 36} (2019), no.~7 075012, [\href{http://arxiv.org/abs/1812.01297}{{\tt arXiv:1812.01297}}].

\bibitem{Romao:2017uwa}
M.~C. Romao and S.~F. King, {\it {Starobinsky-like inflation in no-scale supergravity Wess-Zumino model with Polonyi term}},  {\em JHEP} {\bf 07} (2017) 033, [\href{http://arxiv.org/abs/1703.08333}{{\tt arXiv:1703.08333}}].

\bibitem{Kohri:2013mxa}
K.~Kohri, Y.~Oyama, T.~Sekiguchi, and T.~Takahashi, {\it {Precise Measurements of Primordial Power Spectrum with 21 cm Fluctuations}},  {\em JCAP} {\bf 10} (2013) 065, [\href{http://arxiv.org/abs/1303.1688}{{\tt arXiv:1303.1688}}].

\bibitem{CMB-S4:2016ple}
{\bf CMB-S4} Collaboration, K.~N. Abazajian et~al., {\it {CMB-S4 Science Book, First Edition}},  \href{http://arxiv.org/abs/1610.02743}{{\tt arXiv:1610.02743}}.

\bibitem{Munoz:2016owz}
J.~B. Mu\~noz, E.~D. Kovetz, A.~Raccanelli, M.~Kamionkowski, and J.~Silk, {\it {Towards a measurement of the spectral runnings}},  {\em JCAP} {\bf 05} (2017) 032, [\href{http://arxiv.org/abs/1611.05883}{{\tt arXiv:1611.05883}}].

\bibitem{Modak:2021zgb}
T.~Modak, T.~Plehn, L.~R\"over, and B.~M. Sch\"afer, {\it {Probing the Inflaton Potential with SKA}},  {\em SciPost Phys. Core} {\bf 5} (2022) 037, [\href{http://arxiv.org/abs/2112.09148}{{\tt arXiv:2112.09148}}].

\bibitem{Easther:2021rdg}
R.~Easther, B.~Bahr-Kalus, and D.~Parkinson, {\it {Running primordial perturbations: Inflationary dynamics and observational constraints}},  {\em Phys. Rev. D} {\bf 106} (2022), no.~6 L061301, [\href{http://arxiv.org/abs/2112.10922}{{\tt arXiv:2112.10922}}].

\bibitem{Bahr-Kalus:2022prj}
B.~Bahr-Kalus, D.~Parkinson, and R.~Easther, {\it {Constraining cosmic inflation with observations: Prospects for 2030}},  {\em Mon. Not. Roy. Astron. Soc.} {\bf 520} (2023), no.~2 2405--2416, [\href{http://arxiv.org/abs/2212.04115}{{\tt arXiv:2212.04115}}].

\bibitem{Freedman:2012zz}
D.~Z. Freedman and A.~Van~Proeyen, {\em {Supergravity}}.
\newblock Cambridge Univ. Press, Cambridge, UK, 2012.

\bibitem{Stewart:1994ts}
E.~D. Stewart, {\it {Inflation, supergravity and superstrings}},  {\em Phys. Rev. D} {\bf 51} (1995) 6847--6853, [\href{http://arxiv.org/abs/hep-ph/9405389}{{\tt hep-ph/9405389}}].

\bibitem{Rees:1922}
E.~L. Rees, {\it Graphical discussion of the roots of a quartic equation},  {\em The American Mathematical Monthly} {\bf 29} (1922), no.~2 51--55.

\bibitem{DESI:2024mwx}
{\bf DESI} Collaboration, A.~G. Adame et~al., {\it {DESI 2024 VI: cosmological constraints from the measurements of baryon acoustic oscillations}},  {\em JCAP} {\bf 02} (2025) 021, [\href{http://arxiv.org/abs/2404.03002}{{\tt arXiv:2404.03002}}].

\bibitem{Dimopoulos:2017ged}
K.~Dimopoulos, {\it {Ultra slow-roll inflation demystified}},  {\em Phys. Lett. B} {\bf 775} (2017) 262--265, [\href{http://arxiv.org/abs/1707.05644}{{\tt arXiv:1707.05644}}].

\bibitem{Pattison:2018bct}
C.~Pattison, V.~Vennin, H.~Assadullahi, and D.~Wands, {\it {The attractive behaviour of ultra-slow-roll inflation}},  {\em JCAP} {\bf 08} (2018) 048, [\href{http://arxiv.org/abs/1806.09553}{{\tt arXiv:1806.09553}}].

\bibitem{Bernal:2023wus}
N.~Bernal, S.~Cl{\'e}ry, Y.~Mambrini, and Y.~Xu, {\it {Probing reheating with graviton bremsstrahlung}},  {\em JCAP} {\bf 01} (2024) 065, [\href{http://arxiv.org/abs/2311.12694}{{\tt arXiv:2311.12694}}].

\bibitem{Xu:2024fjl}
Y.~Xu, {\it {Ultra-high frequency gravitational waves from scattering, Bremsstrahlung and decay during reheating}},  {\em JHEP} {\bf 10} (2024) 174, [\href{http://arxiv.org/abs/2407.03256}{{\tt arXiv:2407.03256}}].

\bibitem{Jiang:2024akb}
Y.~Jiang and T.~Suyama, {\it {Spectrum of high-frequency gravitational waves from graviton bremsstrahlung by the decay of inflaton: case with polynomial potential}},  {\em JCAP} {\bf 02} (2025) 041, [\href{http://arxiv.org/abs/2410.11175}{{\tt arXiv:2410.11175}}].

\bibitem{Xu:2025wjq}
X.-J. Xu, Y.~Xu, Q.~Yin, and J.~Zhu, {\it {Full-Spectrum Analysis of Gravitational Wave Production from Inflation to Reheating}},  \href{http://arxiv.org/abs/2505.08868}{{\tt arXiv:2505.08868}}.

\bibitem{Kallosh:2004yh}
R.~Kallosh and A.~D. Linde, {\it {Landscape, the scale of SUSY breaking, and inflation}},  {\em JHEP} {\bf 12} (2004) 004, [\href{http://arxiv.org/abs/hep-th/0411011}{{\tt hep-th/0411011}}].

\end{thebibliography}\endgroup
\end{document}